\documentclass{emulateapj}
\usepackage{apjfonts}
\usepackage{lscape}

\def\ltsima{$\; \buildrel < \over \sim \;$}
\def\simlt{\lower.5ex\hbox{\ltsima}}            
\def\gtsima{$\; \buildrel > \over \sim \;$}
\def\simgt{\lower.5ex\hbox{\gtsima}}            

\newcommand{\asca}{{\it ASCA}}

\newcommand{\chandra}{{\it Chandra}}
\newcommand{\spitzer}{{\it Spitzer}}
\newcommand{\xmm}{{\it XMM-Newton}}
\newcommand{\Xmm}{{\it XMM-NEWTON}}
\newcommand{\subaru}{{Subaru}}
\newcommand{\Subaru}{{SUBARU}}
\newcommand{\scam}{{Suprime-Cam}}
\newcommand{\eexpmap}{{\it eexpmap}}
\newcommand{\eboxdetect}{{\it eboxdetect}}
\newcommand{\emldetect}{{\it emldetect}}
\newcommand{\esplinemap}{{\it esplinemap}}
\newcommand{\esensmap}{{\it esensmap}}

\newcommand{\logn}{log $N$ - log $S$ relation}
\newcommand{\Logn}{Log $N$ - Log $S$ Relation}
\newcommand{\etal}{et al.}

\newcommand{\ergs}{erg cm$^{-2}$ s$^{-1}$}

\newcommand{\de}{deg$^2$}
\newcommand{\nh}{$N_{\rm H}$}
\newcommand{\hun}{$^{\prime}$}
\newcommand{\byo}{$^{{\prime}{\prime}}$}
\newcommand{\cosmo}{($H_0$, $\Omega_{\rm m}$, $\Omega_{\lambda}$)}

\lefthead{Ueda et al.}
\righthead{SXDS (X-ray)}

\begin{document}

\journalinfo{Accepted for publication in ApJS} 
\submitted{}

\title{
The \Subaru /\Xmm\ DEEP SURVEY (SXDS): III. X-RAY DATA}

\author{
Yoshihiro Ueda\altaffilmark{1},
Michael G. Watson\altaffilmark{2},
Ian M. Stewart\altaffilmark{2,3},
Masayuki Akiyama\altaffilmark{4,5},
Axel D. Schwope\altaffilmark{6},
Georg Lamer\altaffilmark{6},
Jacobo Ebrero\altaffilmark{7},
Francisco J. Carrera\altaffilmark{7},
Kazuhiro Sekiguchi\altaffilmark{8},
Tohru Yamada\altaffilmark{5},
Chris Simpson\altaffilmark{9},
G\"unther Hasinger\altaffilmark{10},
Silvia Mateos\altaffilmark{2}
}

\altaffiltext{1}{Department of Astronomy, Kyoto University, Kyoto 606-8502, Japan}
\altaffiltext{2}{Department of Physics \& Astronomy, University of Leicester, Leicester, LE1 7RH, UK}
\altaffiltext{3}{Jodrell Bank Centre for Astrophysics, University of Manchester, Oxford Road, Manchester, M13 9PL, UK}
\altaffiltext{4}{Subaru Telescope,
National Astronomical Observatory of Japan, Hilo, HI, 96720, USA}
\altaffiltext{5}{Astronomical Institute, Tohoku University, Sendai 980-8578, Japan}
\altaffiltext{6}{Astrophysikalisches Institut Potsdam, An der Sternwarte 16, 14482 Potsdam, Germany}
\altaffiltext{7}{Instituto de Fisica de Cantabria (CSIC-UC)
Avda. de los Castros, 39005 Santander, Spain}
\altaffiltext{8}{National Astronomical Observatory of Japan, 2-21-1 Osawa, Mitaka, Tokyo 181-8588, Japan}
\altaffiltext{9}{Astrophysics Research Institute, 
Liverpool John Moores University, Twelve Quays House, Egerton Wharf, Birkenhead CH41 1LD, UK}
\altaffiltext{10}{Max-Planck-Institut f\"ur extraterrestrische Physik,
Giessenbachstrasse 85748 Garching, Germany}

\begin{abstract}

We present the X-ray source catalog in the \subaru /\xmm\ deep
survey. A continuous area of 1.14 \de\ centered at R.A.\ = 02h18m and
Dec.\ = --05d is mapped by seven pointings with \xmm\ covering the
0.2--10 keV band. From the combined images of the EPIC pn and MOS
cameras, we detect 866, 1114, 645, and 136 sources with sensitivity
limits of $6\times10^{-16}$, $8\times10^{-16}$, $3\times10^{-15}$, and
$5\times10^{-15}$ \ergs\ in the 0.5--2, 0.5--4.5, 2--10, and 4.5--10
keV bands, respectively, with detection likelihood $\geq$7
(corresponding to a confidence level of 99.91\%). The catalog
consists of 1245 sources in total including 32 extended-source
candidates. The averaged \logn s are in good agreement with previous
results, bridging the flux range between \chandra\ deep surveys and
brighter surveys. The \logn s show significant spatial variation among
pointings on a scale of 0.2 \de . Analyzing the auto correlation
function, we detect significant clustering signals from the 0.5--2 keV
band sample, which can be fit with a power law form
$(\theta/\theta_c)^{-0.8}$ with a correlation length of
$\theta_c=5.9$\byo $^{+1.0^{{\prime}{\prime}}}_{-0.9^{{\prime}{\prime}}}$ when the integral constraint term
is included. In the 2--10 keV band, however, the clustering is not
significant with a 90\% upper limit of $\theta_c < 1.5$\byo .

\end{abstract}

\keywords{catalogs --- diffuse radiation --- galaxies: active --- X-rays: 
galaxies --- X-rays: general}

\section{Introduction}

The \subaru /\xmm\ Deep Survey (SXDS; \citealt{sek07}) is, along with
COSMOS \citep{sco07,has07}, one of the largest multi-wavelength survey
projects with an unprecedented combination of depth and sky area over
a contiguous region of $>$ 1 \de . The main aims of the SXDS are
to make an accurate measurement of the global properties of the
universe without being affected by cosmic variance and to reveal the
evolution of the large scale structure. The SXDS consists of a wealth
of multi-wavelength data taken by the most modern observing
facilities; X-ray imaging/spectroscopic data in the 0.2--10 keV band
taken by the European Photon Imaging Camera (EPIC;
\citealt{str01,tur01}) onboard \xmm\ \citep{jan01}, multi-color (B, V,
R, i$^\prime$, z$^\prime$) deep optical images by \scam\ on the
\subaru\ telescope \citep{fur07}, deep near infrared maps (J, H, K)
observed as the Ultra Deep Survey (UDS) in the United Kingdom Infrared
Deep Sky Survey (UKIDSS; \citealt{law07}), mid- and far-infrared data
(3.6--160 $\mu$m) taken with the \spitzer\ Space
Observatory\footnote{the \spitzer\ Legacy Survey and part of the
\spitzer\ Wide-area InfraRed Extragalactic (SWIRE) survey}, the
submillimeter (850 $\mu$m) map by the SCUBA Half Degree Extragalactic
Survey (SHADES; \citealt{mor05})\footnote{The SXDS field is also the
field of SCUBA-2 Cosmology Legacy Survey.}, and the deep radio image
(1.4 GHz) by the Very Large Array \citep{sim06}. The SXDS field is
centered at R.A.\ = 02h18m and Dec.\ = --05d and the total field of
the Subaru \scam\ and \xmm\ EPIC images covers an area of 1.3 and 1.14
\de, respectively. The overall survey design and details of each
survey are summarized in \citet{sek07}.

The data of \xmm\ constitute a major component of the SXDS project. X-ray
surveys are a powerful tool to trace the cosmological evolution of
active phenomena in the universe, including Active Galactic Nuclei
(AGNs) and clusters/groups of galaxies. The main constituents of X-ray
sources that make up the X-Ray Background (XRB) are AGNs (for a recent
review, see \citealt{bra05}). Their dominant populations are
``obscured'' AGNs \citep{set89}, where the central engine is blocked by
dust and/or gas in the line of sight. In particular, surveys by hard
X-rays above 2 keV are the most efficient approach to detect these
obscured AGN populations of various luminosity classes with least
bias, thanks to their strong penetrating power against the
photo-electric absorption of matter and small contamination from
stars. In fact, the surface density of AGNs detected in the most
sensitive X-ray surveys \citep[][]{ale03} far exceeds that detected in
the optical bands \citep[e.g.,][]{wol03}, thus providing us with the
most complete and clean samples for AGN studies including heavily
obscured, low luminosity ones that dominate the whole AGN populations by
number.

The tight correlation between the mass of a Super Massive Black Hole
(SMBH) in the galactic center and that of the galactic bulge
\citep{fer00,geb00} indicates strong links between the growth of SMBHs
and star formation in the past. Since AGNs are phenomena that mark the
process of growth of SMBHs in galactic centers, to elucidate the
evolution of AGNs is a fundamental issue for understanding the history
of the universe. Combination of ultra-deep pencil beam surveys and
large area surveys have revealed the cosmological evolution of
the X-ray luminosity function of AGNs
\citep[][]{ued03,laf05,bar05,has05}. The global accretion history of
the universe is similar to that of star formation \citep[see
e.g.,][]{fra99,mar04}. An important finding is that the number density
of more luminous AGNs have a peak at higher redshifts compared with
less luminous ones. This behavior is called ``down-sizing'' or
``anti-hierarchical'' evolution, which is opposite to a naive
expectation from the standard structure formation theory of the
universe. Similar trends were also obtained in the star formation
history \citep[e.g.,][]{cow96,kod04}. These facts imply two modes of
``co-evolution'' of galaxies and SMBHs for different masses. It could
be explained by the feedback from supernova and AGNs
\citep[e.g.,][]{gra04}. Deep multi-wavelength surveys enable us to
observe various populations in different evolutionary stage, giving
the best opportunities to investigate how galaxies and SMBHs
co-evolved with cosmic time.

Despite this major progress on AGN evolution in recent years,
there are several critical issues to be resolved by current and
future X-ray surveys, even if we limit the scope only to X-ray detected
AGNs. Immediate objectives include (1) whether or how the fraction of
obscured AGNs evolves with redshift, (2) the number density of Compton
thick AGNs (those whose line-of-sight absorption exceeds $N_{\rm H} >
10^{24}$ cm$^{-2}$ ), (3) the evolution of the AGN luminosity function
at high redshifts ($z \simgt 4$) to be compared with the results from
optical QSO surveys, and (4) the evolution of clustering properties of
AGNs as a function of luminosity and type.
The SXDS X-ray survey, one of the few ``wide'' and ``deep''
surveys, will give important steps to answer these questions and to
establish the average properties of the AGN populations. Population
synthesis models of the XRB utilize the results from deep surveys in a
limited area at the faintest flux levels \citep{ued03,gil07}, which
would be subject to cosmic variance when we discuss the XRB spectrum
with $\approx$10\% accuracy. This is crucial when we discuss a
relatively minor contribution of some X-ray sources, such as Compton
thick AGNs. To constrain the statistical properties of rare objects,
such as high redshift QSOs, much larger cosmic volume is necessary than
in the currently available deep surveys. By surveying a continuous area
with a large depth, we can also study the evolution of the large scale
structure probed by AGNs. It helps us to understand the physical
conditions that triggers AGN activity for different luminosity class,
type (1 or 2), and redshift. At the flux limits of the SXDS, we can
detect the most dominant populations of X-ray sources that contribute to
the XRB. In all of these studies, high completeness of identification
(redshift determination) is crucial. The multi-wavelength data are
particularly useful in identifying X-ray sources even if optical
spectroscopy is difficult for a population of ``optically faint'' AGNs
\citep{ale01}.

In this paper, we describe the X-ray data of the SXDS based on the \xmm\
survey performed in 2000 and 2003. We present the whole X-ray catalog,
and basic statistical properties of the detected X-ray sources. The
X-ray data have been used in a number of studies in the SXDS projects,
detection of ultra luminous X-ray sources in nearby galaxies
\citep{wat05}, mid-IR and radio selected Compton-thick AGNs
\citep{mar07}, radio sources \citep{sim06}, Lyman $\alpha$ sources
\citep{sai07,ouc07}, and optical variability selected AGNs
\citep{mor07}. A series of papers on optical identification of the X-ray
sources are forthcoming (Akiyama \etal , in preparation).
The paper is organized as follows. \S~2 summarizes the observation of
the \xmm . In \S~3 we describe the detailed procedure of data
analysis, and present the source list and its statistics. \S~4 and
\S~5 shows the results of \logn s and auto correlation functions,
respectively. In \S~6 long-term time variability of X-ray sources is
studied in a selected field. \S~7 summarizes the
conclusions. Throughout the paper, the cosmological parameters \cosmo\
= (100$h$ km s$^{-1}$ Mpc$^{-1}$, 0.3, 0.7) are adopted.

\section{Observations}

The SXDS field, centered at R.A.\ = 02h18m and Dec.\ = --05d, was
mapped with \xmm\ in the 0.2--10 keV band by seven pointings, one deep
(nominal exposure 100 ksec) observation in the center surrounded by
six shallower (50 ksec each) ones. \xmm\ carries three EPICs, pn,
MOS1, and MOS2, each has a field of view (FOV) of about
$30'\times30'$. The pointing positions are arranged so that the
combined X-ray map, having a ``flower petal'' pattern, covers
essentially the whole region of the optical images taken with the
\subaru\ telescope, mosaic of 5 \scam\ images in each band
\citep{fur07}. There are small overlapped regions between these
pointings, making the X-ray image continuous without gaps over an area
of about 1.14 \de , although the sensitivity is not completely uniform
over the entire map due to different exposures and instrumental
effects, such as vignetting (see \S~4.1 for details).

Table~1 gives the observation log, including the pointing position,
observation time, and net exposure obtained after screening out high
background time. We designate the pointings as SDS-1 for the central
field and SDS-2 through SDS-7 for the surrounding ones in a clockwise
direction. The observations were performed in 3 discrete epochs, the
first from 2000 July 31 to August 8 (for SDS-1, 2, 3, 4), the second
from 2002 August 8 to 12 (for SDS-5, 6, 7), and the third on 2003
January 7 to supplement the unfulfilled exposure of SDS-4. The
observation of each pointing were performed continuously except for
short intervals (several hours) in SDS-1 and SDS-3 and a long interval
(two and half years) in SDS-4. The ``thin'' filters were used for the
three cameras in all the observations.

\begin{deluxetable*}{rrrrrrrr}
\tabletypesize{\footnotesize}
\tablenum{1}
\tablecaption{Log of the \xmm\ Observations in the SXDS field\label{tbl-1}} 
\tablehead{
\colhead{Field}&\colhead{Obs.\ ID}&
\colhead{RA\tablenotemark{a}}&\colhead{Dec\tablenotemark{a}}&\colhead{PA\tablenotemark{b}}&
\colhead{Start}& \colhead{End} &\colhead{Exposure\tablenotemark{c}}\\
\colhead{}&\colhead{}&
\colhead{(deg)}&\colhead{(deg)}&\colhead{(deg)}&
\colhead{(UT)}& \colhead{(UT)} &\colhead{(sec)}}
\startdata
SDS-1
&0112370101 &34.47838 &-4.98117 &73.6 &2000/07/31 22:38 &08/01 14:04 &40361 \\
&0112371001 &34.47883 &-4.98050 &74.5 &2000/08/02 21:13 &08/03 13:15 &42473 \\

SDS-2
&0112370301 &34.87900 &-4.98047 &74.8 &2000/08/04 21:05 &08/05 14:00 &40220 \\

SDS-3
&0112370401 &34.67971 &-4.63422 &75.7 &2000/08/06 07:53 &08/06/17:38 &14341 \\
&0112371501 &34.67958 &-4.63414 &75.7 &2000/08/06 20:58 &08/06/23:20 &4080 \\

SDS-4
&0112371701 &34.27946 &-4.63394 &76.0 &2000/08/08 07:57 &08/08 15:02 &21277 \\
&0112372001 &34.32404 &-4.66858 &245.7&2003/01/07 04:41 &01/07 11:56 &25940 \\

SDS-5
&0112370601 &34.07854 &-4.98164 &76.1 &2002/08/12 06:06 &08/12 18:52 &34377 \\

SDS-6
&0112370701 &34.27754 &-5.32797 &74.9 &2002/08/08 15:28 &08/09 04:43 &46802 \\

SDS-7
&0112370801 &34.67792 &-5.32856 &74.9 &2002/08/09 05:52 &08/09 19:02 &36803 
\enddata
\tablenotetext{a}{Mean pointing position (J2000) of the optical axis.}
\tablenotetext{b}{Mean position angle}
\tablenotetext{c}{Net exposure for the pn data after screening}
\end{deluxetable*}

\section{Analysis}

\subsection{Image Production}

The reduction of the EPIC data was performed by using the Science
Analysis System (SAS) software. We used the SAS version
6.1.0\footnote{The latest Current Calibration Files (CCF) as of 2005
January 31 were used.} for image creation and all the subsequent
analysis, while an earlier version of SAS (that very similar to
version 5.3.3) was utilized to produce the event files from which we
start the image analysis, processed in the \xmm\ Survey Science Center
(SSC) Pipeline Processing System (PPS) facility in University of
Leicester.

As the first step, we created sky images with a bin size of 4 arcsec
in four energy bands, 0.3--0.5, 0.5--2, 2--4.5, 4.5--10 keV,
separately for each observation ID and detector. The 7.9--8.3 keV band
was excluded to avoid the instrumental background of Cu K-lines
\citep{str01}. With the standard event selection, we accumulated good
photon events with ``patterns'' of 0 through 12 (MOS1 and MOS2) or 0
through 4 (pn). The time region of background flare was excluded by
examining the light curve. This leaves a net exposure of about
60--80\% of the allocated observing time, as listed in Table~1.

To achieve the best positional accuracy, we corrected the absolute
astrometry of the EPIC image of each obs.\ ID with respect to the
\subaru\ R-band image in the following way. We performed source
detection from each observation, and cross correlated the obtained
(tentative) source list with the optical catalog \citep{fur07} to
calculate the mean positional offsets. Then we corrected the original
images for these offsets, which were found to be 0\byo -- 1.3\byo ,
consistent with the estimated error in the absolute astrometry of \xmm . 
This correction also minimizes systematic errors caused by summing
up multiple observations for the case of SDS-1, 3 and 4.

\subsection{Source Detection}

We produce source lists separately from different pointings (SDS-1
through 7). Later, these are merged into one list by excluding
overlapping regions to define a statistically independent sample from
the whole data (\S~3.3).  The overall flow of the source detection is
similar to the one adopted in the SSC PPS used to produce the 2XMM
catalog (Watson \etal , in preparation), although several improvements
are employed here. The main differences are as follows; (1) we perform
source detection to the summed image of pn, MOS1, and MOS2, not to
individual ones. (2) We optimize several parameters of the spline
interpolation in making background maps. (3) Special care is paid to
separate neighboring sources.

\subsubsection{Background Maps}

For each observation, energy band, and detector, we create a
background map from the image by spline interpolation with the SAS
program \esplinemap\ after excluding sources found with \eboxdetect ,
which perform simple source detection based on a cell detection
algorithm. After optimization of parameters, we verify that the
obtained models well represent the profiles of the background (i.e.,
the unresolved XRB plus non X-ray background) over the whole FOV
without large deviation from the data.

\subsubsection{Summation of Images}

We sum up the pn, MOS1, MOS2 images and background maps for each
energy band and each ``pointing'', which means also combining those of
multiple observations for SDS-1, 3 and 4. This enables us to achieve
the best sensitivity from the whole available data in the simplest
manner. At the same time, we can avoid technical problems in the
position and flux determination that would be caused by image gaps
between CCD chips in a single detector.

To obtain exposure maps, we first calculated them for each instrument
in narrow energy bands of every 0.5 keV, using the SAS task \eexpmap .
Then, we took their count-rate weighted average, normalized to the pn
count rate in a given survey band. The count rate distribution is
calculated through the energy response of each instrument assuming a
power law photon index of 1.5 above 2 keV and 2.0 below 2 keV. The
dependence on assumed photon indices is negligibly small in the
analysis. The cross-calibration of effective area between the pn and
MOS cameras is known to be accurate within 5\% level, and its
uncertainty does not affect the results.

With this procedure, we are able to treat the summed images and
exposure maps as if they were created from a single pn detector
regardless of the detected position.  In the following analysis, we
present the flux of a source in terms of a ``pn-equivalent'' count
rate. Strictly speaking, this treatment may not be perfectly accurate
in the image fitting process described below, because the positions of
the optical axis in the combined image are not common for different
detectors and observations, which could affect a precise modeling of
the combined Point Spread Function (PSF) of the mirrors.  To estimate
{\it maximum} systematic errors caused by this effect, we make the
same analysis by changing the position of the optical axis to that of
a MOS camera. We find that the fluxes of sources obtained in the two
analyses match each other by 1.3\% level, confirming that this
approximation is justified at this accuracy.

Figure~1 shows the pseudo three-colored X-ray image of the SXDS field
combined from the 7 pointings. This image is made by the {\it export}
command in the IRAF package\footnote{IRAF is distributed by the
National Optical Astronomy Observatories, which is operated by the
Association of Universities for Research in Astronomy, Inc.\ under
cooperative agreement with the National Science Foundation}. As the
inputs, we use the exposure-corrected, background subtracted
pn+MOS1+MOS2 images in the 0.5--2, 2--4.5, and 4.5--10 keV bands,
corresponding to red, green, and blue colors, respectively. They are
smoothed by a Gaussian profile with a 1$\sigma$ width of 4 arcsec over
the entire field. Larger color fluctuation is apparently seen in outer
regions compared with the central region due to the shorter exposure.

\begin{figure*}
\epsscale{1.0}
\plotone{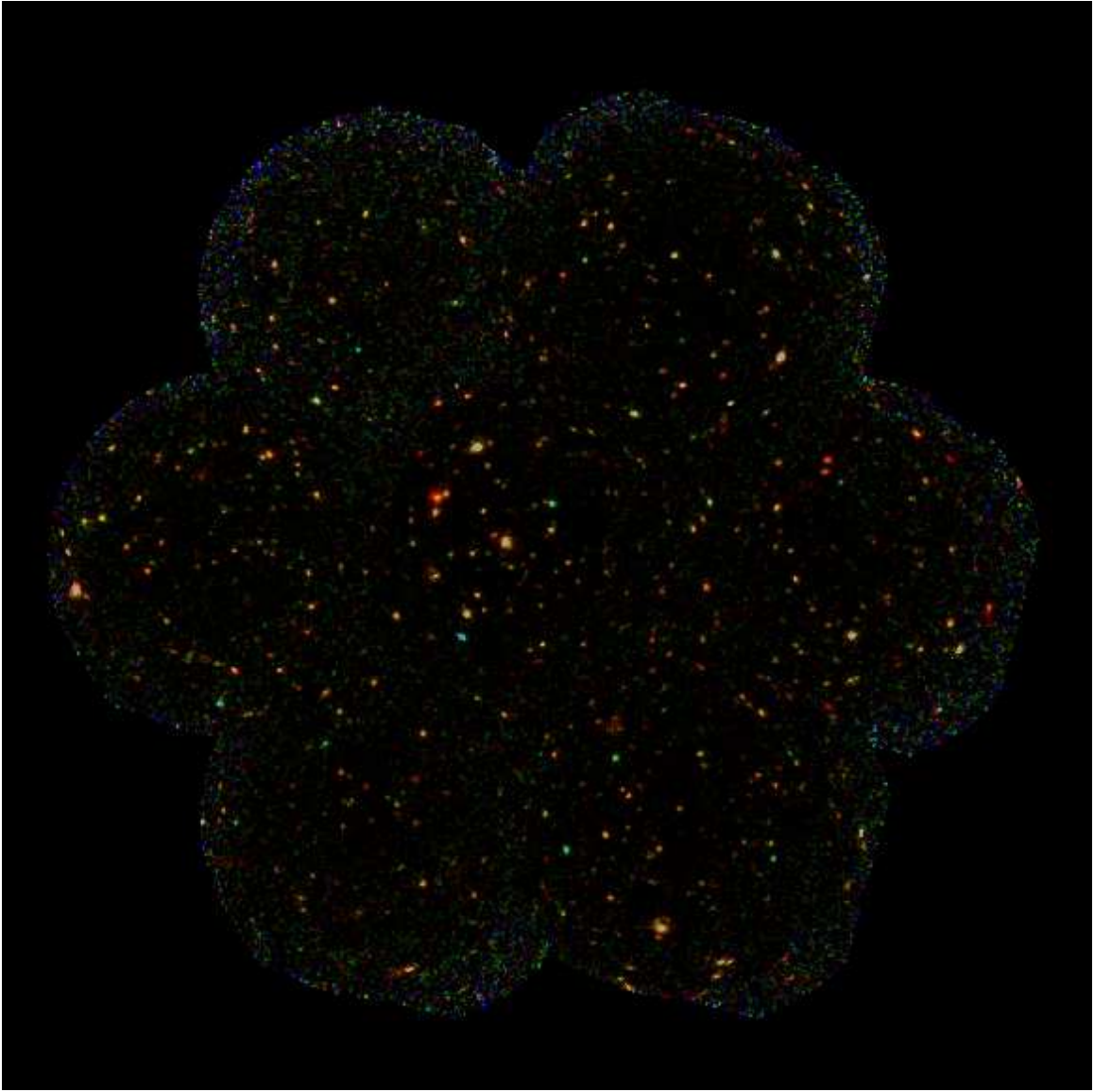}
\caption{The smoothed, 3-colored X-ray image of the whole SXDS field 
obtained from the energy bands of 0.5--2 keV (red), 2--4.5 keV
 (green), and 4.5--10 keV (blue).
\label{fig1}}
\end{figure*}

\subsubsection{Maximum Likelihood Fit}\label{sec:MLF}

The source list from each pointing is obtained by the maximum
likelihood fit applied to the detector co-added images in the four
energy bands by using the SAS task \emldetect . The \emldetect\
program reads images, background maps, and exposure maps in a single
or multiple energy bands, and makes a simultaneous fit to the images
with a model consisting of PSFs over the background map, based on the
input list of source candidates. It returns the fitted position,
vignetting-corrected count rate, and detection likelihood (hereafter
called ``ML'' standing for Maximum Likelihood) of each source. The
obtained count rate corresponds to the total flux in the entire PSF.

To make the list complete, this process is iterated twice in the
following manner. We first perform cell detection by \eboxdetect\ with
a cell size of 20\byo $\times$ 20\byo\ to obtain a list of source
candidates with minimum likelihood of 3.2, which is used as an input
to the first run of \emldetect .  In both programs we allow the flux
and position of a source to be free parameters. We find that in some
cases the above cell size is too large to completely detect sources in
a high density region. To supplement this, we also perform
cell-detection with a 12\byo $\times$ 12\byo cell size, and add new
source candidates detected with high significance into the source list
obtained above. In this stage, by human inspection, we pay attention
not to include obviously fake sources such as a part of a PSF tail
from nearby bright sources, diffuse emission, and false detection
close to gaps between CCD chips. The merged list is again input to
\emldetect , producing a final source list.

Through the fitting process, we find that there are sources that are
likely to be confused by the PSF of neighboring ones in crowded
regions. To measure the fluxes of these sources with the best
accuracy, we perform an image fit in each region by allowing the
positions and fluxes of the multiple sources (up to three) to vary
simultaneously in order to solve the coupling of the fluxes
self-consistently. Their angular separation is typically 12\byo --
24\byo. Their fluxes are replaced with the new values obtained here,
and the ML values are calculated for each source at the fixed
position. The number of pairs of such possibly confused sources are
about 10 per pointing, and are marked by flag ``C'' in the last column
of the source catalog (Table~2).

The statistic that \emldetect\ uses to fit the source parameters is
the $C$ statistic defined by \citet{cas79}. For a Poissonian
probability distribution, appropriate to the measurement of X-ray
events by \xmm , Cash's statistic takes the form
\begin{displaymath}
 	C = 2 \sum_{i=1}^{N} \sum_{j=1}^{M}  (e_{i,j} - n_{i,j} 
\ln{e_{i,j}}) \ + \ \textrm{a constant,}
\end{displaymath}
where $e_{i,j}$ is the expected number of X-ray events in the $i$th
pixel and $j$th energy band, $n_{i,j}$ is the detected number of
events in that pixel/band, and the sum is over the $N$ pixels within
the detection region and the $M$ energy bands used (here = 4 as
stated). The value of $e_{i,j}$ is obtained by adding the value of the
source model at that pixel/band to the expected contribution from the
background; $e_{i,j}$ is thus a function of the source model
parameters.

After arriving at those values of the source parameters which minimize 
$C$, the detection likelihood (formally, the probability of the null 
hypothesis) for those optimum values is then calculated. Cash's 
prescription for this is to form the difference
\begin{displaymath}
 	\Delta C = C_\mathrm{null} - C_\mathrm{best}
\end{displaymath} 
where $C_\mathrm{null}$ is $C$ calculated by setting the amplitude of the 
source model to zero and $C_\mathrm{best}$ is the minimum result returned 
by the fitting routine. According to Cash's theory, $\Delta C$ is 
distributed approximately as $\chi^2$ for $\nu$ degrees of freedom, where 
$\nu$ is the number of fitted parameters. The probability $p(\chi^2 \ge 
\Delta C)$ of obtaining the calculated value of $\Delta C$ or greater by 
chance fluctuations of the detected background can therefore be obtained 
in terms of the complementary incomplete gamma function $Q$ as follows:
\begin{displaymath}
 	p(\chi^2 \ge \Delta C) = Q(\nu/2,\Delta C/2).
\end{displaymath}
The value of ML is set within \emldetect\ to equal $-{\rm ln}(p)$.
Note that the statistics must be treated with caution when the total
number of photon counts used for the fit are very small ($\simlt$ 9),
which is not the case in our analysis.

The ML value as calculated in the above way is viewed as more
sensitive than the simple box-detection statistic used by \eboxdetect
, because it uses information about the source PSF to help reject
random fluctuations of the detected background. Offsetting this is a
possible bias introduced by the fact that \emldetect\ is not given
'free rein' over the whole field, but is applied only to a small
number of restricted areas of the field, at locations of candidate
sources already found by \eboxdetect . If the detection threshold of
\eboxdetect\ is set to too high a value, there is a risk that a real
source with a good shape, but not many counts, will be missed by the
combined detection process. In order to avoid this bias, one ought
therefore to run the preceding \eboxdetect\ with a detection threshold
set deliberately low. On the other hand, since \emldetect\ must go
though the computationally-intensive fitting process with each of its
candidates, it will be impractical to provide it with too long a list
of candidates.  The respective values of 3.2 and 7 adopted here for
the \eboxdetect\ and \emldetect\ likelihood cutoffs were chosen with
this necessary compromise in mind.

An additional advantage of using \emldetect\ to determine the final source 
parameters is its ability to add the models of already fitted sources to 
an internal background map. Since the sources are processed in the order 
of their brightness, it is possible to take into account the background 
introduced by bright sources when fitting the fainter sources.

\subsubsection{Examining Source Extent}

Up to this stage, all sources are assumed to be point like. To
constrain the spatial extent of the detected sources, we perform an
image fit by allowing the source extent to be a free parameter
assuming a Gaussian profile. The source positions are fixed at the
input ones. This procedure yields a list of extended source candidates
from the SXDS field (Table~3). For these sources, the fluxes listed in
Table~2 correspond to those obtained in this process with
consideration of the source extent. However, as our procedure is
essentially dedicated to detection of point like sources in the
earlier stage, this list must be regarded to be incomplete. A more
extensive approach to search for extended sources will be presented
elsewhere (Finoguenov \etal , in preparation).

\begin{deluxetable}{lcc}
\tabletypesize{\footnotesize}
\tablenum{3}
\tablecaption{The list of Candidates of Extended Sources in the SXDS\label{tbl-3}}
\tablehead{
\colhead{No.}&\colhead{Extent Likelihood}& \colhead{Extent\tablenotemark{a}} \\
\colhead{}& \colhead{}& \colhead{($''$)}}
\startdata
0034&  5.1 & 2.8$\pm$ 0.5 \\
0051& 11.4 &17.9$\pm$ 1.2 \\
0140&  6.8 &22.0$\pm$ 3.3 \\
0153&  9.2 & 4.7$\pm$ 0.4 \\
0156&  7.4 & 4.2$\pm$ 0.9 \\
0239&  4.4 & 3.9$\pm$ 1.0 \\
0280&  4.5 &19.3$\pm$ 3.3 \\
0285& 36.6 &16.0$\pm$ 1.8 \\
0287&  4.9 & 4.8$\pm$ 1.3 \\
0292&  5.5 & 4.1$\pm$ 1.3 \\
0396&  4.0 & 5.9$\pm$ 1.8 \\
0441& 12.4 &12.7$\pm$ 1.4 \\
0453&  7.2 &12.4$\pm$ 2.2 \\
0514& 30.9 &21.0$\pm$ 2.0 \\
0552&  6.5 & 7.1$\pm$ 1.4 \\
0621&  6.5 & 6.6$\pm$ 1.4 \\
0622&  4.3 & 4.6$\pm$ 1.4 \\
0625&  5.7 & 2.9$\pm$ 1.8 \\
0646&  9.1 & 3.5$\pm$ 0.8 \\
0647& 10.3 & 9.4$\pm$ 1.6 \\
0712&  6.6 & 4.3$\pm$ 0.8 \\
0784& 14.3 &33.6$\pm$ 3.3 \\
0796& 20.5 & 3.0$\pm$ 0.3 \\
0829& 48.6 &11.5$\pm$ 0.6 \\
0852&  4.7 & 3.3$\pm$ 0.2 \\
0876& 19.6 &17.4$\pm$ 1.6 \\
0889&  5.7 &17.2$\pm$ 3.1 \\
0934& 10.0 & 4.7$\pm$ 0.4 \\
1152&  5.0 &11.8$\pm$ 3.9 \\
1168&  4.6 &18.4$\pm$ 2.8 \\
1169&  5.2 & 3.7$\pm$ 1.1 \\
1176& 31.4 &15.3$\pm$ 1.1 

\enddata
\tablenotetext{a}{The source extent when a Gaussian profile is adopted ($1\sigma$).}
\end{deluxetable}

\subsection{Source List}

Thus, we obtain seven source lists separately from different pointings
(SDS-1 through 7), containing the information of position and fluxes
with the ML values in the four energy bands, 0.3--0.5, 0.5--2, 2--4.5,
and 4.5--10 keV. In these tentative lists, we keep all sources whose
summed ML from the four bands is equal to or exceeds 5, which will be
further screened in the following way.

To make scientific research in comparison with previous studies, it is
quite useful to define a sample selected in standard bands such as
0.5--4.5 keV (hereafter ``XID band'') and 2--10 keV band (``hard
band''). Although the combined flux and ML can be calculated from
those in the two individual bands (i.e., 0.5--2 and 2--4.5 keV for the
XID band, and 2--4.5 and 4.5--10 keV for the hard band), its
statistical treatment would become quite complex; for instance, the
sensitivity limit cannot be uniquely determined as a function of ML at
a given position, depending on the source spectrum. Hence, to obtain a
well defined sample selected in the XID or hard band in the same way
as in the four narrow bands, we calculate their ML values by repeating
the likelihood fit using the 0.5--4.5 or 2--10 keV image with the
background map summed from the two narrow bands. In the fit, the
positions of the sources are fixed\footnote{Since we fix the position
in the XID and hard bands, the corresponding $\Delta C$ value is
smaller than those obtained in the four individual bands for a given
ML, because the degree of freedom $\nu$ is reduced by 2. This is
also the case for the ``confused'' sources whose ML values are obtained
by fixing the position (see \S~\ref{sec:MLF}). We properly take these
effects into account in the analysis of \logn s and auto correlation
function.} and no new sources are considered.

In this paper we adopt a threshold for the detection likelihood of 7
in a single band. Thus, from the tentative source lists, we select
only those detected with ML$\geq$7 either in the 0.3--0.5 keV
(ultra-soft), 0.5--2 keV (soft), 2--4.5 keV (medium), 4.5--10 keV
(ultra-hard), 0.5--4.5 keV (XID), or 2--10 keV (hard) band, to be
included in the final list. Further, for statistical analysis using
this sample, it is convenient to have a single source list merged from
the seven pointings; as mentioned above, there are overlapping regions
between different pointings in the outer edge of the FOV, where same
sources are repeatedly detected. To exclude such duplication, we only
refer to the results of a single pointing that achieves the highest
sensitivity at a given position, based on sensitivity maps in the XID
band (see below).

Table~2 gives the source list in the SXDS field complied in this way,
sorted by right ascension (RA) and declination (Dec): (col.\ 1) source
number; (cols.\ 2 and 3) the X-ray source position as determined by
\xmm\ (RA and Dec); (col.\ 4) the statistical error
in the position estimated through the maximum likelihood fit (root sum
square of the 1$\sigma$ error in each direction); (cols.\ 5 through 10)
the ML value in each energy band; (cols.\ 11 through 14) the
vignetting-corrected count rate in each energy band; (cols.\ 15 trough
17) the hardness ratios defined as $HR1=(S-US)/(S+US)$,
$HR2=(M-S)/(M+S)$, $HR3=(UH-M)/(UH+M)$, where $US$, $S$, $M$, and $UH$
are the count rates in the 0.3--0.5, 0.5--2, 2--4.5, 4.5--10 keV band,
respectively; (col.\ 18) the pointing ID at which the source is
detected; (col.\ 19) the offset angle from the mean optical axis in
the corresponding pointing; (col.\ 20) the total pn-equivalent
exposure (sum of pn, MOS1, and MOS2) at the source position, corrected
for the vignetting in the 0.5-4.5 keV band; (col.\ 21) the background
rate in the 0.5-4.5 keV band at the source position; (col.\ 22) the
flag if the flux is determined by multiple-source fit to take into
account possible source confusion with nearby sources (see \S~3.2.3).
The exposure and background rate (cols.\ 20 and 21) are averaged over
the source extraction region with a fixed encircled energy fraction
(68\%) of the PSF. From the results of optical identification (Akiyama
et al., in preparation), we confirm that the positional errors given
in this list are reasonable. In Table~3 we list the spatial extent and
its likelihood of all the extended sources contained in Table~2.

We detect 1245 sources in total. The numbers of detected sources in a
single or two bands in any combination from the six energy bands are
summarized in Table~4. According to simulation study \footnote{see
http://xmmssc-www.star.le.ac.uk/newpages/1XMM/fig\_5.3.html }, the
number of fake sources due to statistical fluctuation is estimated to
be about 1.5 per pointing in each energy band at our likelihood
threshold (ML$\geq$7), hence $\approx 50$ in the sum sample. Their
fraction in the total number of detected sources depends on the
selection band; it becomes the smallest (0.6\%) in the 0.5--4.5 keV
but largest (5\%) in the 4.5--10 keV band because of different
sensitivities.

\begin{deluxetable*}{ccccccccc}
\tablenum{4}
\tablecaption{Number of Detected Sources\label{tbl-4}} 
\tablehead{
\colhead{Detection Band}& 
\colhead{0.3--0.5} &\colhead{0.5--2} & \colhead{2--4.5}& \colhead{4.5--10}&
\colhead{0.5--4.5} &\colhead{2--10} & 
\colhead{One Band Only\tablenotemark{a}}}
\startdata
0.3--0.5 keV& {\bf 357}& 339& 235& 73& 341& 253& 16\\
0.5--2 keV&      & {\bf 866}& 412& 113& 853& 466& 13\\
2--4.5 keV &      &    & {\bf 487}& 121& 474& 461& 4\\
4.5--10 keV&     &    &    & {\bf 136}& 125& 133& 3\\
0.5--4.5 keV&    &    &    &    & {\bf 1114}& 550& 169\\
2--10    keV&    &    &    &    &     & {\bf 645}& 78
\enddata
\tablenotetext{a}{Number of sources detected only in one band given in the first column}
\tablecomments{Number of sources commonly detected in the combination of two bands {\bf (or a single band)} are listed.}
\end{deluxetable*}

We calculate the conversion factors from a count rate into a flux in
each energy band by using the energy response of the pn. Galactic
absorption of \nh\ = $2.5\times10^{20}$ cm$^{-2}$ \citep{dic90} is
taken into account. The values are summarized in Table~5 for a power
law spectrum with various photon indices ($\Gamma$). Figure~2(a) shows
their dependence on the photon index assuming no absorption, whereas
in Figure~2(b) we change the absorption column density at zero
redshift for a fixed photon index of 1.8. In the same figures we also
plot the hardness ratios $HR1$, $HR2$, and $HR3$ as a function of
spectral parameters. For the flux conversion, hereafter we assume
$\Gamma=1.5$ (with no absorption) for the 2--4.5 keV, 4.5--10 keV,
0.5--4.5 keV, and 2--10 keV band, and $\Gamma=2.0$ for the 0.3--0.5
and 0.5--2 keV band, unless otherwise mentioned.

\begin{deluxetable*}{ccccccc}
\tabletypesize{\footnotesize}
\tablenum{5}
\tablecaption{Flux Conversion Factors\label{tbl-5}}
\tablehead{
\colhead{Assumed Spectrum}&
\multicolumn{6}{c}{Flux Conversion Factors\tablenotemark{a}}\\
\colhead{Photon Index}& \colhead{0.3--0.5 keV}& \colhead{0.5--2 keV}& \colhead{2--4.5 keV}& \colhead{4.5--10 keV\tablenotemark{b}}& \colhead{0.5--4.5 keV}& \colhead{2--10 keV\tablenotemark{b}}}
\startdata
1.0& 0.164& 0.171& 0.574& 1.718& 0.305& 1.057 \\
1.1& 0.167& 0.171& 0.571& 1.696& 0.295& 1.026 \\
1.2& 0.171& 0.170& 0.568& 1.675& 0.285& 0.996 \\
1.3& 0.174& 0.170& 0.564& 1.653& 0.276& 0.967 \\
1.4& 0.177& 0.169& 0.561& 1.632& 0.267& 0.939 \\
1.5& 0.181& 0.168& 0.558& 1.612& 0.259& 0.912 \\
1.6& 0.184& 0.168& 0.555& 1.593& 0.250& 0.886 \\
1.7& 0.187& 0.167& 0.551& 1.573& 0.243& 0.862 \\
1.8& 0.190& 0.167& 0.548& 1.554& 0.236& 0.838 \\
1.9& 0.194& 0.167& 0.545& 1.536& 0.229& 0.816 \\
2.0& 0.197& 0.166& 0.541& 1.517& 0.223& 0.794 
\enddata
\tablenotetext{a}{
The conversion factor from the vignetting-corrected, pn-equivalent count rate 
into energy flux between the same band in units of [10$^{-14}$ \ergs ] / [count ksec$^{-1}$]. The fluxes are corrected for Galactic absorption of $N_{\rm H}=2.5\times10^{20}$ cm$^{-2}$.}
\tablenotetext{b}{
The 7.9--8.3 keV band is excluded in the count rate but included
in the flux.
} 
\end{deluxetable*}

\begin{figure*}
\epsscale{1.0}
\plotone{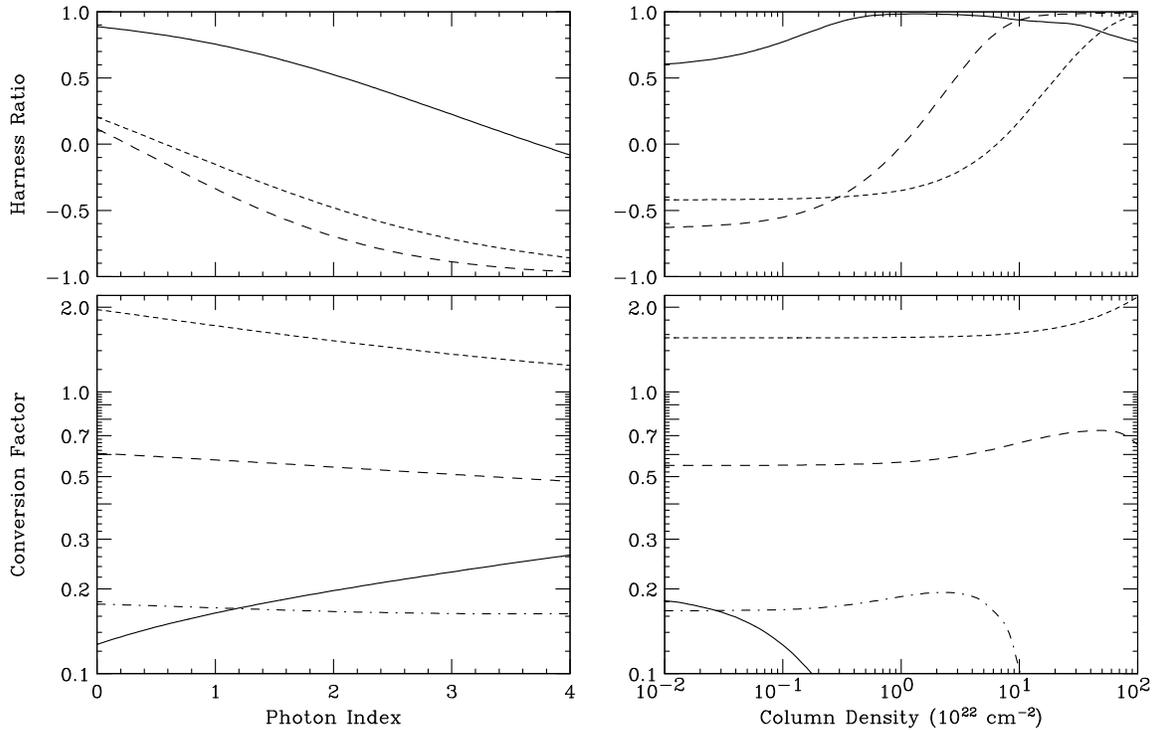}
\caption{({\it Lower}): The conversion factors from the vignetting-corrected count rate into flux in the same band, calculated for different spectra. The
unit is [$10^{-14}$ \ergs\ / (c/ksec)]. The solid, dot-dashed, 
long-dashed, short-dashed curves correspond to the 0.3--0.5 keV, 0.5--2 keV, 2--4.5 keV, and 4.5--10 keV bands, respectively. 
({\it Upper}): The hardness ratio expected from the assumed spectra. The solid, long-dashed, and short-dashed curves correspond to $HR1$, $HR2$, 
and $HR3$, respectively. \newline
(a) {\it Left}: the parameter is a photon index of a power law with no absorption.
(b) {\it Right}: the parameter is an absorption column density for a power law spectrum with a photon index of 1.8.
\label{fig2}}
\end{figure*}

\subsection{Spectral Properties of the Sources}

In this subsection, we summarize spectral properties of the SXDS
sources based on hardness ratio analysis using the above source
list. Figures 3(a) shows the flux distribution of sources detected in
the 0.5--4.5 and 2--10 keV bands. As described below, our sample
covers X-ray sources that constitute a major fraction of the XRB below
10 keV. Figure~3(b) shows the histogram of hardness ratio $HR2$ and
$HR3$ for those detected in the 0.5--4.5 and 2--10 keV bands,
respectively. The flux versus hardness ratio plots are given in
Figures 4 (0.5--4.5 keV flux versus $HR2$) and 5 (2--10 keV flux
versus $HR3$). For clarity we do not attach error bars to each point
in these figures; the typical $1\sigma$ errors in $HR2$ and $HR3$ are
0.22 and 0.15 at flux levels of $(2-4)\times10^{-15}$ \ergs\ (0.5--4.5
keV) and $(1-2)\times10^{-14}$ \ergs\ (2--10 keV), respectively.  To
examine the mean spectral properties as a function of flux, we produce
flux-weighted spectra in the four bands from flux-range limited
samples, by summing up the count rate of the individual sources in 
each energy band. Table~6 summarizes their $HR2$
and $HR3$ values with corresponding power-law photon indices. We find
evidence that the spectra of the X-ray sources becomes harder at
fainter fluxes with a confidence level of 99.99\% ($HR2$) and 96\%
($HR3$) for the 0.5--4.5 keV sample but with a lower confidence level
of 84\% ($HR2$) and 90\% ($HR3$) for the 2--10 keV sample. Their mean
slope is similar or even harder than that of the XRB, $\Gamma \simeq
1.4$ over the 0.5--10 keV band at these flux levels. This result is
consistent with the fact that we already resolved a major parts of the
XRB and that at brighter fluxes where 20--30\% of the XRB is produced,
the mean spectra are much softer than the XRB spectrum \citep{ued99}.

\begin{deluxetable*}{cccccccc}
\tabletypesize{\footnotesize}
\tablenum{6}
\tablecaption{Average Hardness Ratio for Flux Limited Samples\label{tbl-6}} 
\tablehead{
\multicolumn{3}{c}{Sample}&\colhead{}&
\multicolumn{4}{c}{Averange Spectra\tablenotemark{a}}\\
\cline{1-3} \cline{5-8}\\
\colhead{Detection}& \colhead{Flux Range} &\colhead{Number of}&\colhead{}&
\multicolumn{2}{c}{0.5--4.5 keV}&
\multicolumn{2}{c}{2--10 keV}\\
\colhead{Band}&\colhead{($10^{-14}$ erg s$^{-1}$ cm$^{-2}$)}&\colhead{Sources}&\colhead{}&
\colhead{$<HR2>$}&\colhead{$<\Gamma_{0.5-4.5}>$}&\colhead{$<HR3>$}&\colhead{$<\Gamma_{2-10}>$}}
\startdata
0.5--4.5 keV& 0.5--2.0&406&\colhead{}&
$-0.55\pm0.01$& $1.55\pm0.02$&$-0.27\pm0.02$& $1.34\pm0.05$\\
0.5--4.5 keV& 2.0--8.0&107&\colhead{}&
$-0.61\pm0.01$& $1.71\pm0.02$&$-0.33\pm0.02$& $1.52\pm0.05$\\
2--10 keV& 1.0--4.0&334&\colhead{}&
$-0.53\pm0.01$& $1.50\pm0.02$&$-0.23\pm0.02$& $1.23\pm0.04$\\
2--10 keV& 4.0--16&38&\colhead{}&
$-0.55\pm0.01$& $1.55\pm0.02$&$-0.29\pm0.03$& $1.40\pm0.08$
\enddata
\tablenotetext{a}{The averaged hardness ratio is derived the summed spectrum 
of the flux-limited sample. The corresponding power law photon index in the 
same band is also shown, which is corrected for Galactic absorption of $N_{\rm H}=2.5\times10^{20}$ cm$^{-2}$. The error is $1\sigma$.}
\end{deluxetable*}

\begin{figure*}
\epsscale{1.0}
\plotone{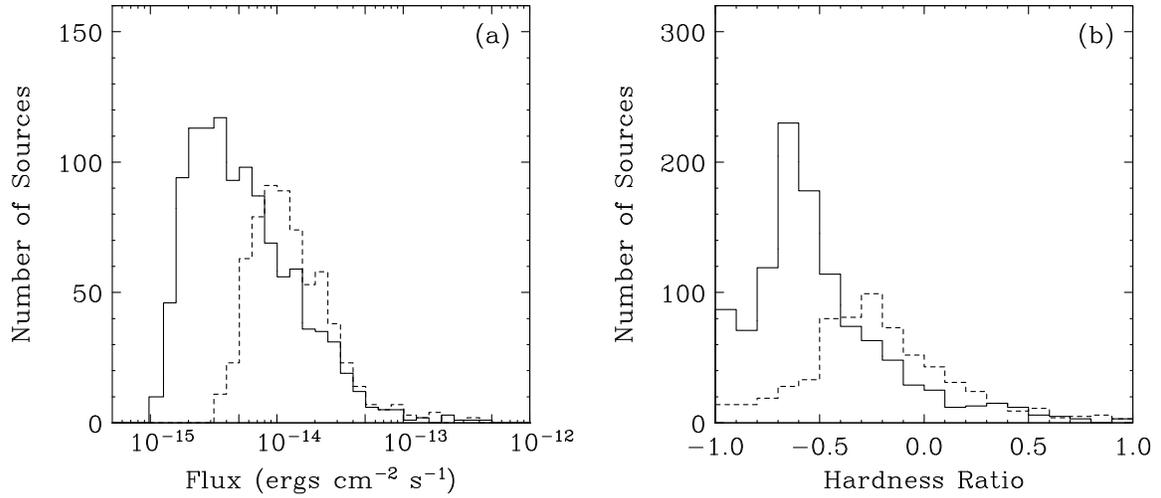}
\caption{
{\it Left:} histogram of flux for the 0.5--4.5 keV (solid) and
 2--10 keV (dashed) selected sample in the SXDS \xmm\ catalog. 
{\it Right:} that of hardness ratio $HR2$ (solid) and $HR3$ (dashed) 
for the 0.5--4.5 keV and 2--10 keV selected sample, respectively.
\label{fig3}}
\end{figure*}

\begin{figure}
\epsscale{1.0}
\plotone{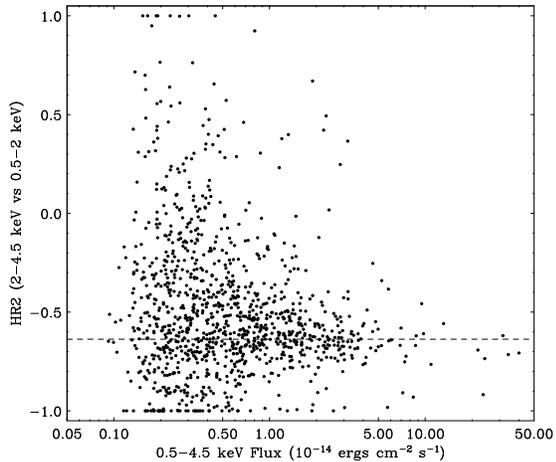}
\caption{
The 0.5--4.5 keV flux vs $HR2$ plot for the sources detected
in the 0.5--4.5 keV band. $HR2$ is defined as $(M-S)/(M+S)$ where
$M$ and $S$ is the vignetting-corrected pn-equivalent count rate in the
 2--4.5 keV and 0.5--2 keV, respectively. The dashed line corresponds
to a power law photon index of 1.8.
\label{fig4}}
\end{figure}

\begin{figure}
\epsscale{1.0}
\plotone{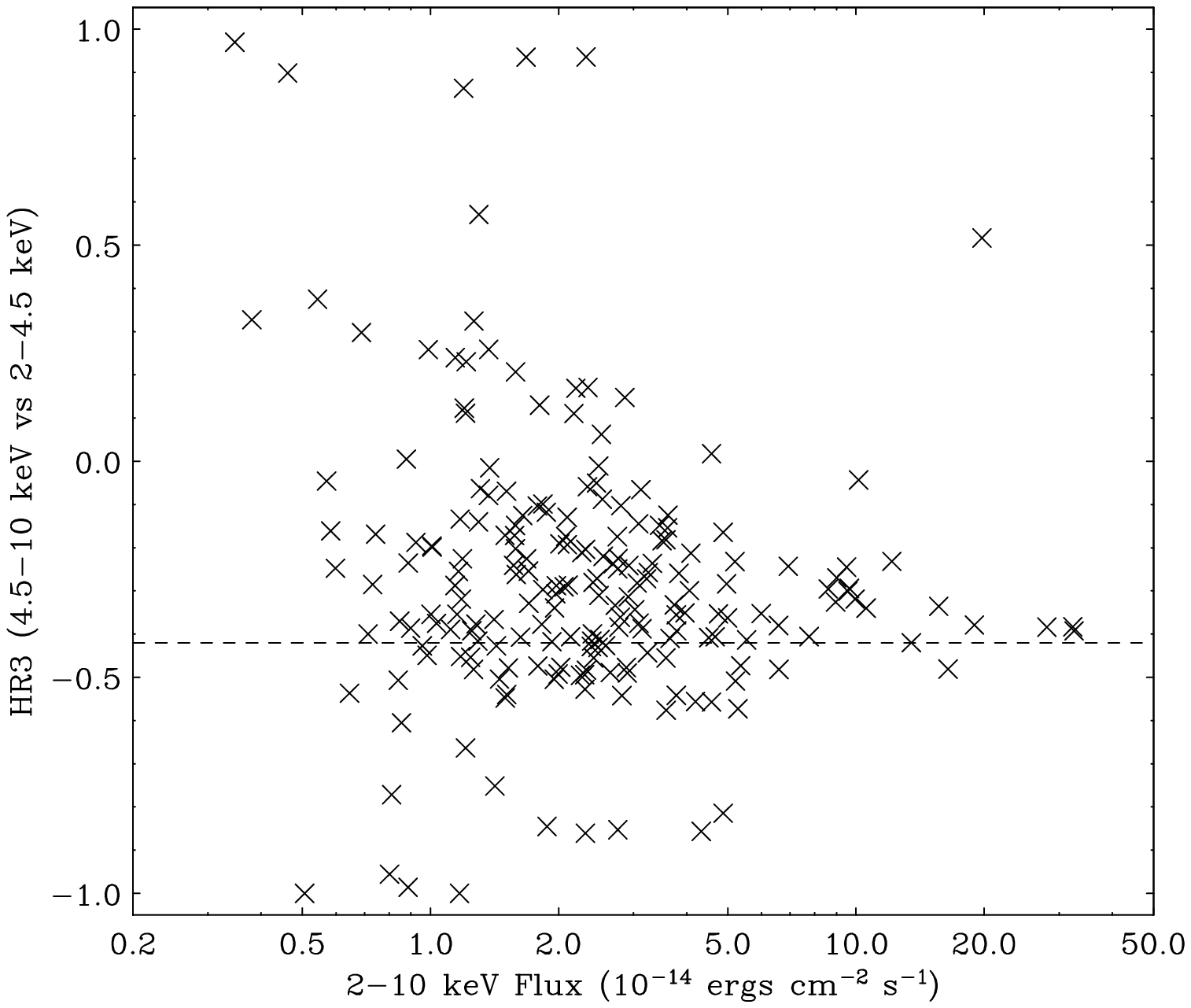}
\caption{The 2--10 keV flux vs $HR3$ plot for the sources detected in
the 2--10 keV band. $HR3$ is defined as $(H-M)/(H+M)$ where $H$ and
$M$ is the vignetting-corrected pn-equivalent count rate in the 4.5-10
keV and 2--4.5 keV, respectively. The dashed line corresponds
to a power law photon index of 1.8.
\label{fig5}}
\end{figure}

Figure~6 shows the color-color plot between $HR2$ and $HR3$ by using
208 sources detected in the 0.5--4.5 and/or 2--10 band whose
statistical errors in both hardness ratios are smaller than 0.2. The
solid curve tracks the color of a power law spectrum with varying
photon indices from 0 to 4, while the dashed curve does that
with varying absorption column densities at $z=1$ (a typical redshift
of our sample; see Figure~14) from Log \nh\ = 0 to 24 for a fixed
photon index of 1.8. As noticed, the spectra of some sources are not
simply represented by an absorbed power law model. In particular,
there is a population of sources that are hard in the 2--10 keV band
(e.g., $HR3\simgt0$) but soft in the 0.5--4.5 keV band ($HR2\simlt0$),
indicating the presence of additional soft components. Detailed
spectral analysis using information of redshift will be presented in
forthcoming papers.

\begin{figure}
\epsscale{1.0}
\plotone{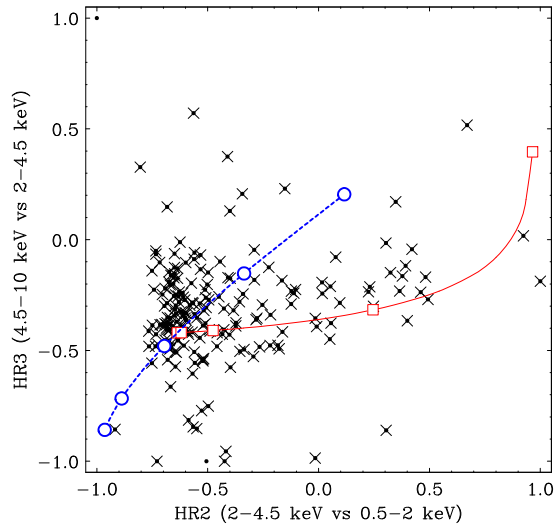}
\caption{
The color-color plot between $HR2$ and $HR3$. Only sources with
a 1$\sigma$ uncertainty less than 0.2 in both $HR2$ and $HR3$ are
included in the plot. The solid curve (red) tracks the colors of an absorbed
power law spectrum for a fixed photon index of 1.8 with varying column
densities (from log \nh\ cm$^{-2}$ = 0, 21, 22, 23, 24 at $z=1$), 
while the dashed one (blue) for varying photon indices from $0$ to
$4$ with no absorption.
\label{fig6}}
\end{figure}

\section{\Logn }

\subsection{Sensitivity Map}

To make statistical analysis using the source list, such as
determination of \logn, it is crucial to have reliable sensitivity
maps for each pointing and energy band, i.e., we need to know a flux
(count rate) limit as a function of position at a given detection
criterion (i.e., ML $\geq 7$). Since we utilize the maximum likelihood
algorithm, however, it is not trivial to calculate sensitivities by
employing an analytic formula, unlike in the case of cell
detection. Ideally, only a detailed simulation can give the correct
estimate of a sensitivity at every position, since it depends on the
background, exposure map, and PSF in a complex way. To save computing
time, we here take an empirical approach by utilizing both simulation
and analytical calculation, as described below.

First, we directly estimate the flux limit by {\bf simulation} at
2-dimensional grid points with 40\byo\ spacing over the FOV. For a
given position, we produce a simulated image where a point source is
superimposed on an input background map, using the \emldetect\
program. Since we are looking simply for a relation between the
emldetect ML value and the flux of the sources, we do not taken into
account Poisson counting noise in our simulation. We then perform a
likelihood fit with \emldetect\ to the simulated image and derive the
ML value for the input source. Repeating the simulation assuming
several different fluxes for the point source, we obtain an empirical
relation between the input flux and output ML value. This relation
enables us to calculate the sensitivity for a given threshold of ML at
that position.

Next, we produce full resolution (4\byo\ pixel) sensitivity maps by {\bf
interpolating} between the grid points in the following manner. The
flux limit could change quite sensitively with the position, being
affected by data gaps between CCD gaps or hot pixel regions. A simple
interpolation of the flux limit based on the rough position sampling
in the above simulation may not be sufficiently accurate. To estimate
the precise position dependence, we utilize analytical sensitivity
maps calculated by the \esensmap\ program, which is based on the cell
detection algorithm. Generally, the detection likelihood obtained by
\emldetect\ (ML) differs from that defined in the cell detection
(ML$_{\rm cell}$) for the same flux. The ratio between ML and ML$_{\rm
cell}$ should be a function of position, primarily depending on the
size and shape of the PSF. Hence, we calculate the ML$_{\rm cell}$
value corresponding to the flux limit at each grid point, and
interpolate them between the grid points as a linear function of
position. We finally convert ML$_{\rm cell}$ into a ``flux limit'' at
each position, by referring to a set of sensitivity maps calculated by
\esensmap\ with different thresholds of ML$_{\rm cell}$, thus
producing the final sensitivity maps. To evaluate systematic errors in
the sensitivity map, we compare the results when the positions of the
grid points are shifted by 20\byo\ in each direction. We find that the
difference in the flux limit is mostly within a few percent and
$<$10\% at maximum. We regard this (10\%) as the maximum relative
systematic error in the sensitivity map, which will be considered in
the analysis of the spatial distribution of sources (\S~\ref{sec:ACF}).

\subsection{Results and Comparison with Other Work}

Superposing the sensitivity maps of the seven pointings in the common
sky coordinates, we examine, position by position, in which pointing
data the best sensitivity is achieved in the 0.5--4.5 keV band in
overlapping regions. As mentioned above, we only refer to the results
of a single pointing, at given sky coordinates, to obtain the combined
source list and sensitivity maps to eliminate duplication. 
The resultant sensitivity map in the 0.5--4.5 keV band is shown in Figure~7.
Figure~8 shows the area curves $\Omega (S)$ for ML$\geq$7 (i.e., survey area
where the detection completeness is guaranteed above the given flux)
obtained from the combined SXDS field in the four energy bands, 0.5--2
keV, 0.5--4.5 keV, 2--10 keV, and 4.5--10 keV.

\begin{figure}
\epsscale{1.4}
\plotone{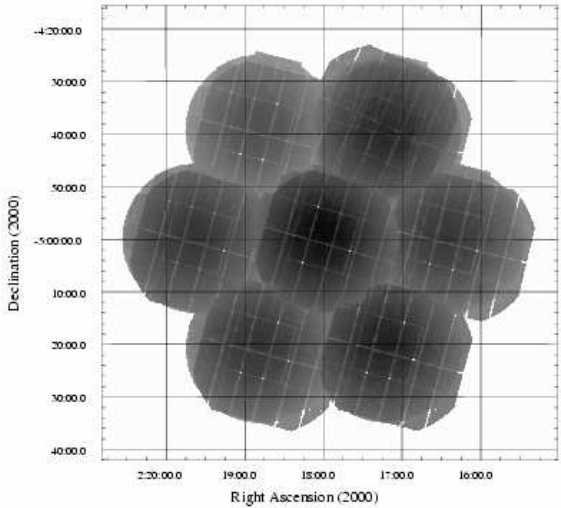}
\caption{
The sensitivity map in the 0.5--4.5 keV band. The minimum
flux threshold for the detection likelihood $\geq 7$ is given as a function
of position in the sense that a darker color corresponds to a lower flux.
SDS-2 is located to the east of the central field, SDS-1, 
and SDS-3 through 7 go in a clockwise direction after SDS-2.
\label{fig7}}
\end{figure}

\begin{figure*}
\epsscale{0.39}
\plotone{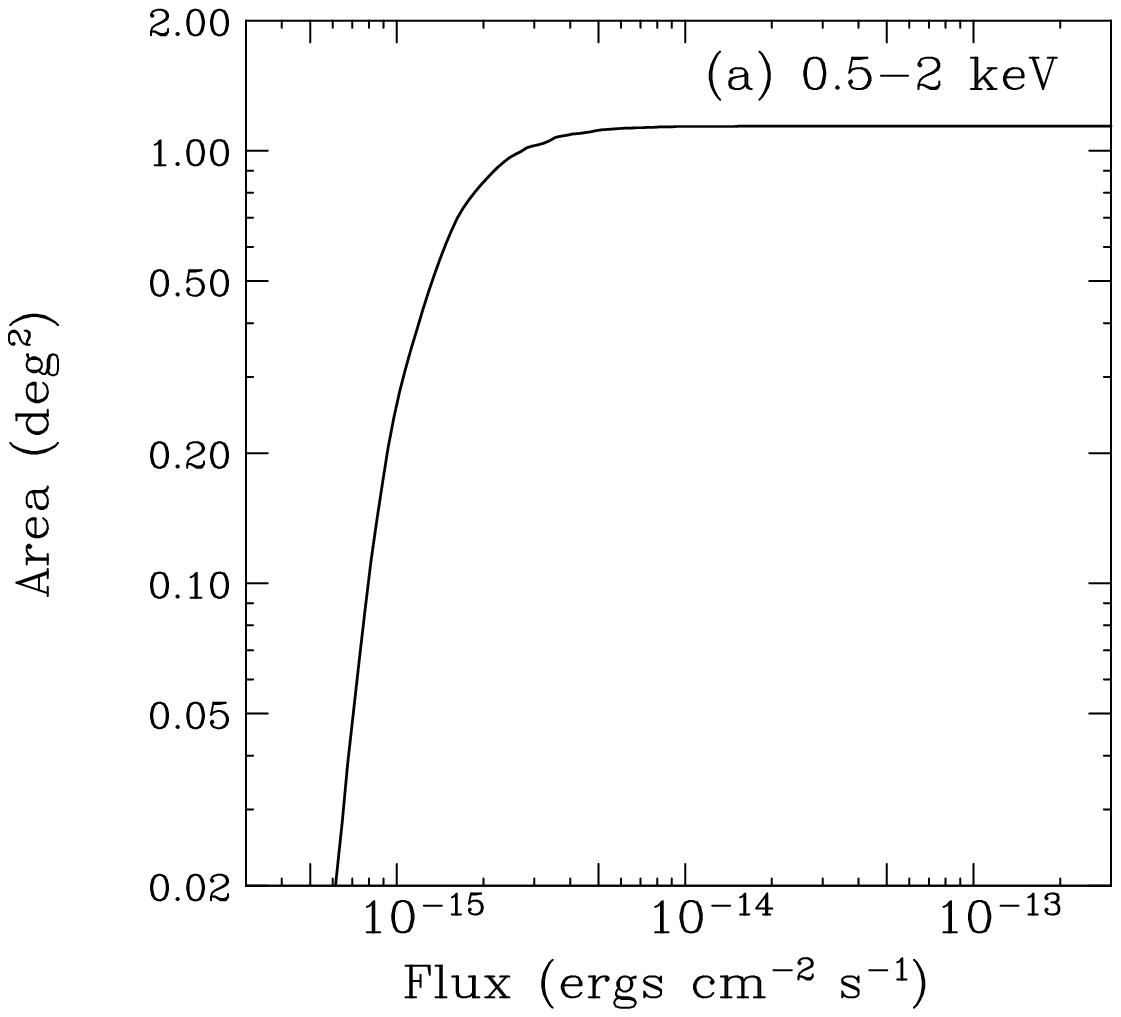}
\plotone{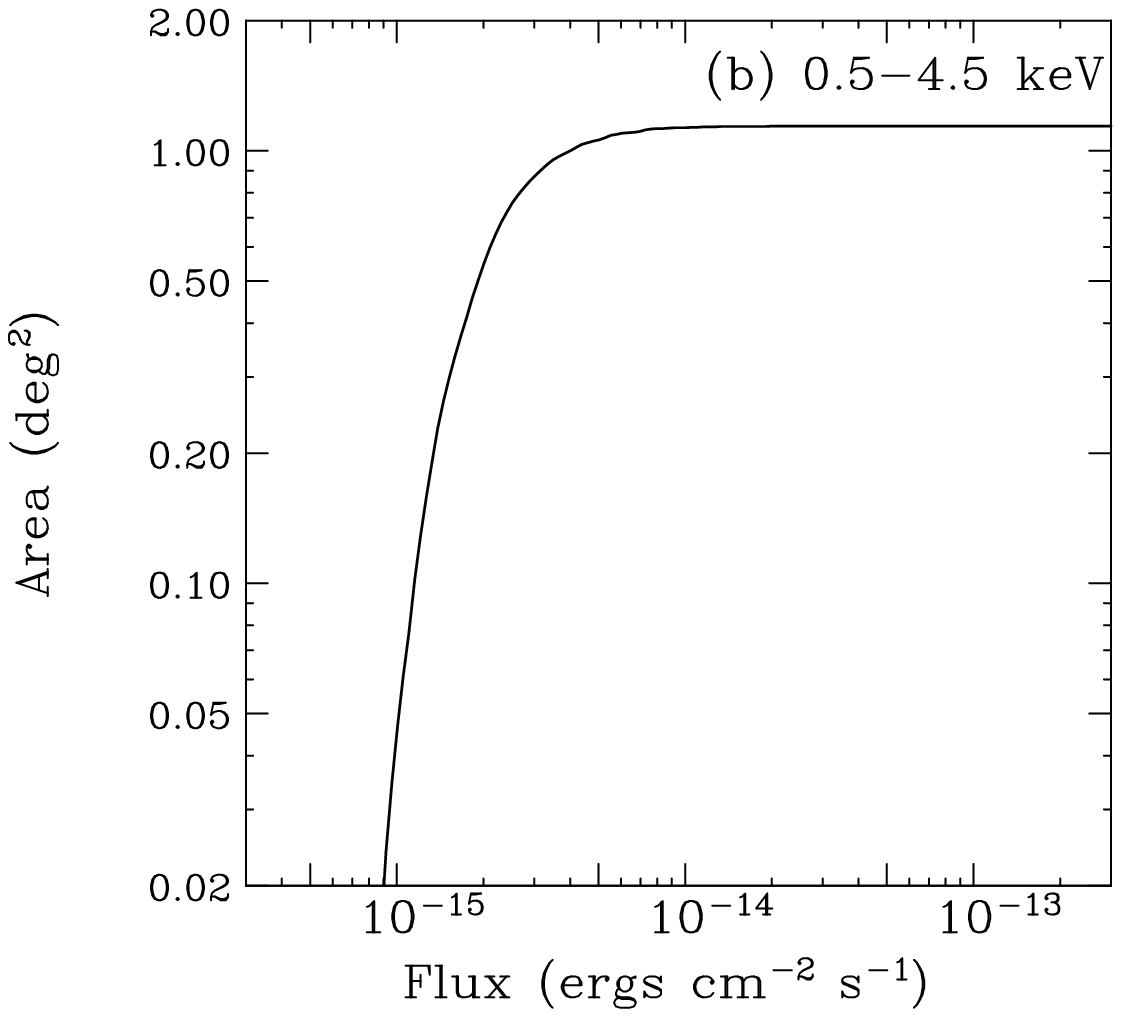}
\plotone{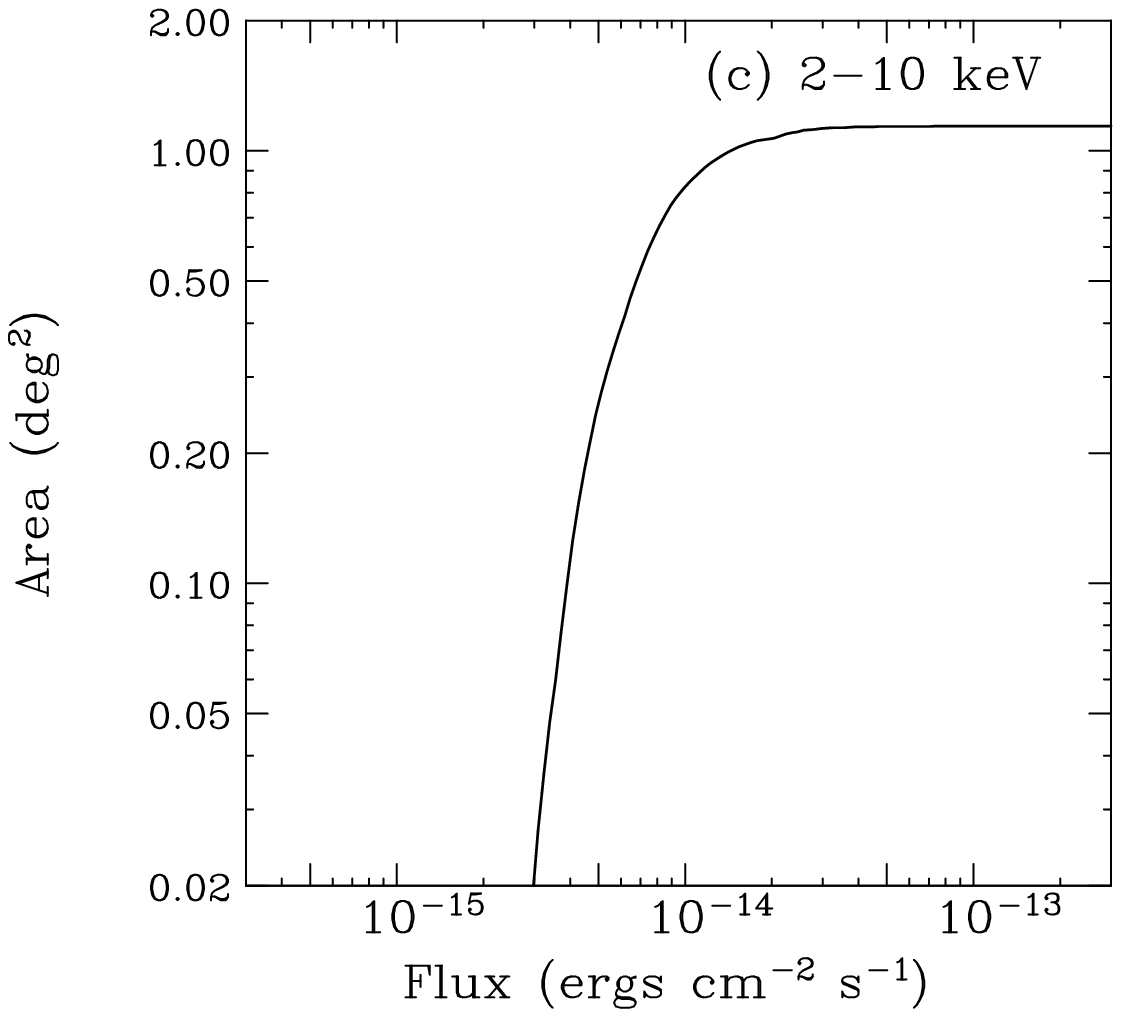}
\plotone{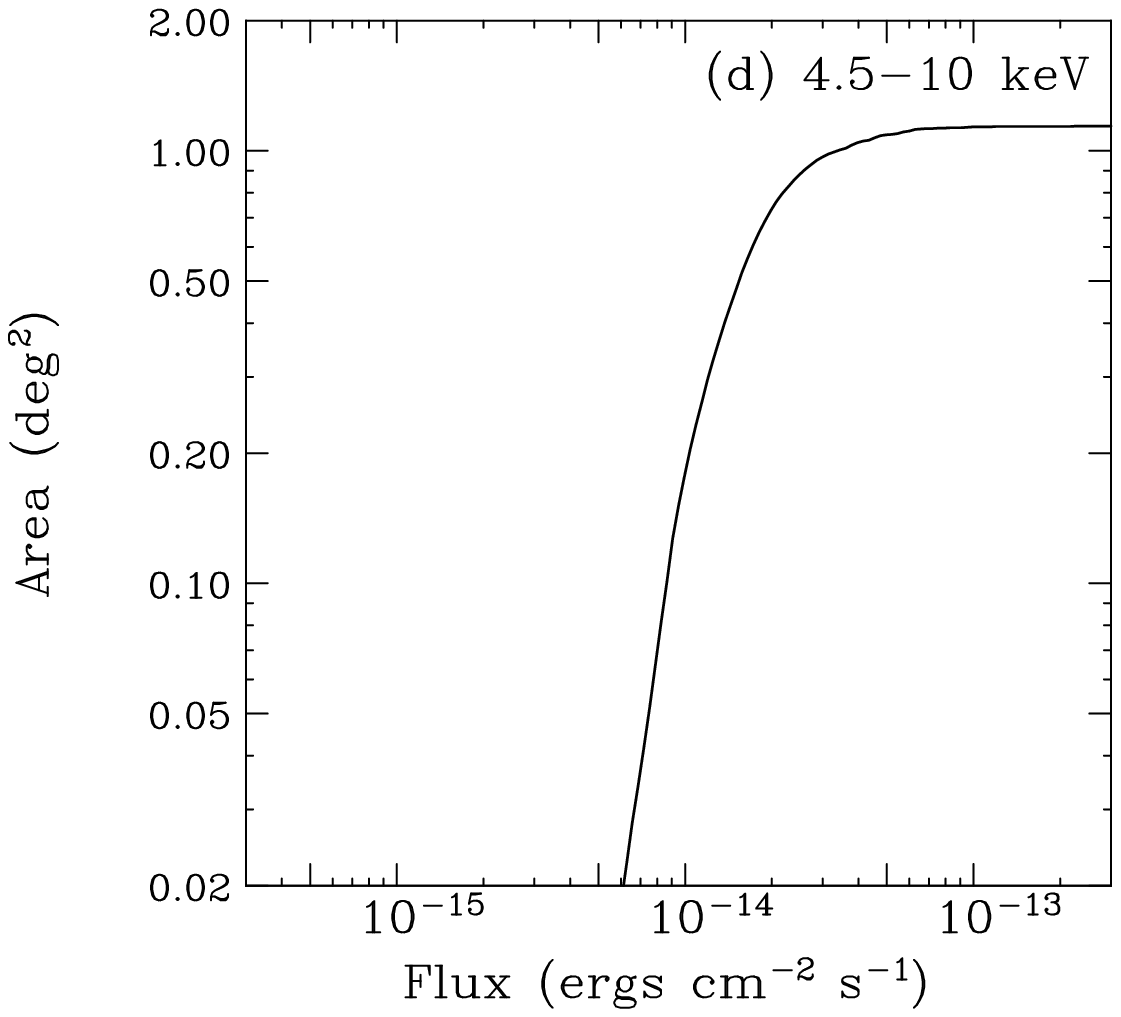}
\caption{
The survey area given as a function of flux in the whole SXDS
field in the (a) 0.5--2, (b) 0.5--4.5, (c) 2--10, and (d) 4.5--10 keV band.
\label{fig8}}
\end{figure*}

Dividing the observed flux distribution by $\Omega (S)$, we first
obtain the \logn s in the differential form, $N(S)$. Here, we discard
sources whose flux falls below the sensitivity limit at the give 
position because we consider that such detection may not be
reliable. Figure~9 shows $N(S)$ in the four bands in units of number per
square degree per $10^{-14}$ \ergs, where the errors represent the 
$1\sigma$ Poisson error in the number of sources in each flux bin.
Following previous work, we fit them with a broken power law form,
expressed as
\begin{equation}\label{eq:diff}
\begin{tabular}{lll}
&$\frac{K}{S_{\rm b}} (\frac{S}{S_{\rm b}})^{-\Gamma_{\rm d}}$ & $(S \leq S_{\rm b})$ \\
$N(S)$ = &&\\
&$\frac{K}{S_{\rm b}} (\frac{S}{S_{\rm b}})^{-\Gamma_{\rm u}}$ & $(S > S_{\rm b})$.
\end{tabular}
\end{equation}
The model has a break flux $S_{\rm b}$, above and below which the
slope is $\Gamma_{u}$ and $\Gamma_{d}$, respectively. To estimate the
parameters, we utilize a Maximum Likelihood method, so that the model
best reproduces the observed flux distribution expected from the area
curve. We fix $\Gamma_{\rm u}$ at 2.5 in the 0.5--2 keV, 0.5--4.5 keV,
and 2--10 keV bands, while in the 4.5--10 keV band, we adopt
a single power law by setting $\Gamma_{d} = \Gamma_{u}$ and
$ S_{\rm b}=1\times10^{-14}$ \ergs\ because the break is not evident in
our flux range. The best-fit models are shown in solid lines in
Figure~9. Table~7 summarizes the best-fit parameters, which are found
to be consistent with previous results within uncertainties
\citep[e.g.,][]{car07,bru08}.

\begin{deluxetable}{ccccc}
\tablenum{7}
\tablecaption{Best Fit Parameters for \Logn s\label{tbl-7}} 
\tablehead{
\colhead{Band}& \colhead{$\Gamma_{\rm u}$} & \colhead{$\Gamma_{\rm d}$} & 
\colhead{$S_{\rm b}$} & \colhead{$K$} \\
\colhead{}& \colhead{}&\colhead{}&
\colhead{($10^{-14}$ cgs)}&\colhead{(deg$^{-2}$)}}
\startdata
0.5--2 keV& 2.5 (fixed) & $1.63{^{+0.07}_{-0.12}}$ &
$1.03{^{+0.25}_{-0.29}}$ & $154{^{+63}_{-32}}$ \\
0.5--4.5 keV& 2.5 (fixed) & $1.63{^{+0.06}_{-0.09}}$ &
$1.81{^{+0.44}_{-0.40}}$ & $177{^{+52}_{-36}}$ \\
2--10 keV& 2.5 (fixed) & $1.41{^{+0.17}_{-0.20}}$ &
$1.19{^{+0.13}_{-0.15}}$ & $444{^{+66}_{-58}}$ \\
4.5--10 keV& $2.62{^{+0.13}_{-0.12}}$ & \nodata &
1.0 (fixed) & $288{^{+40}_{-36}}$
\enddata
\tablecomments{The error is 1$\sigma$ for a single parameter.}
\end{deluxetable}

\begin{figure*}
\epsscale{0.39}
\plotone{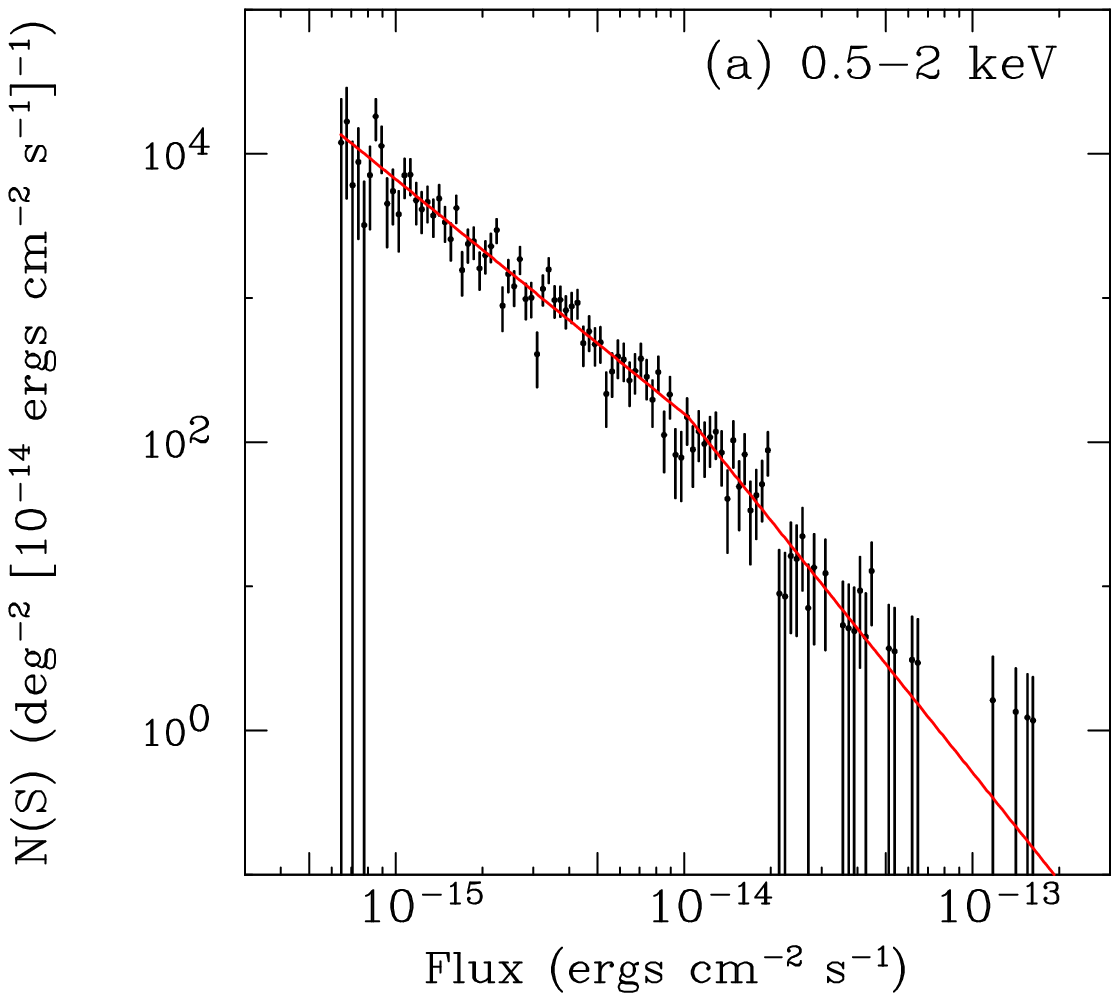}
\plotone{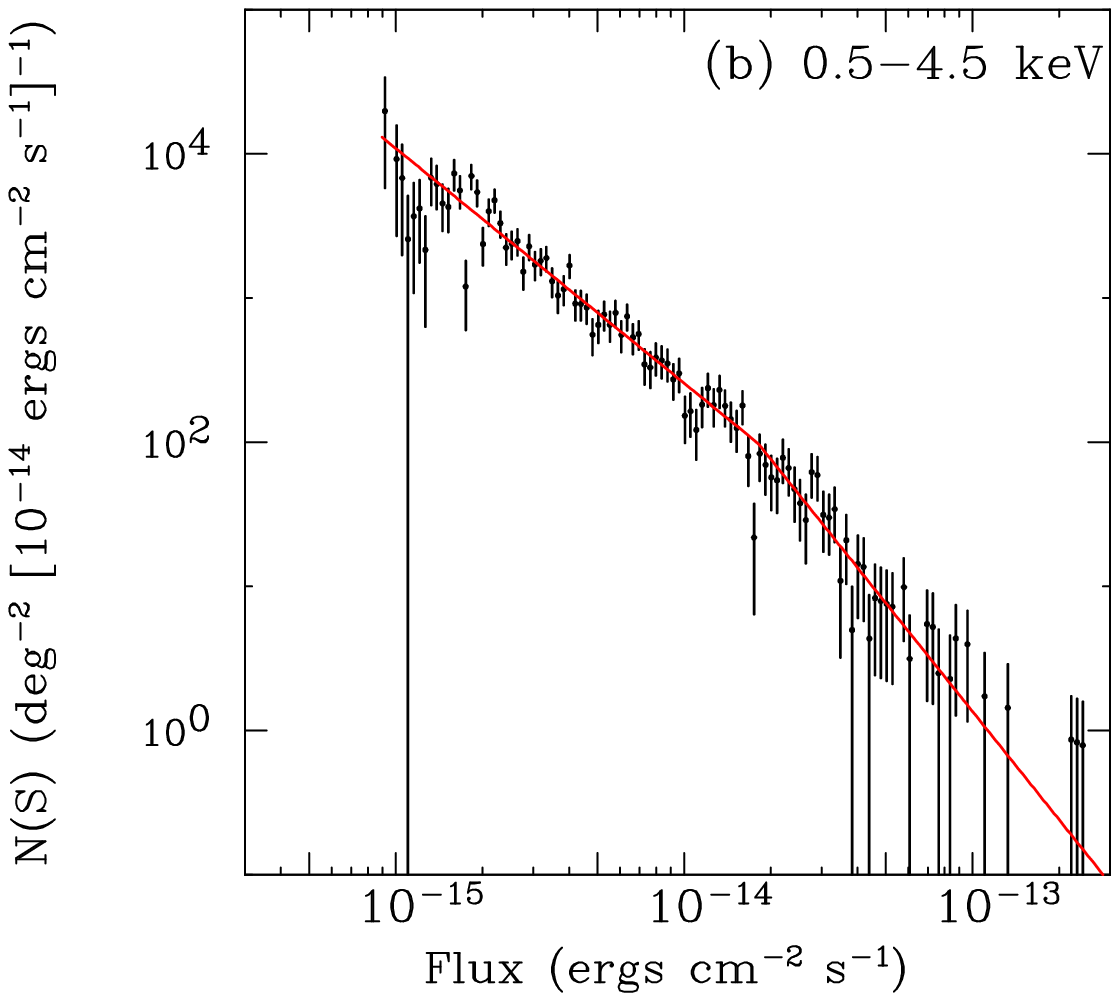}
\plotone{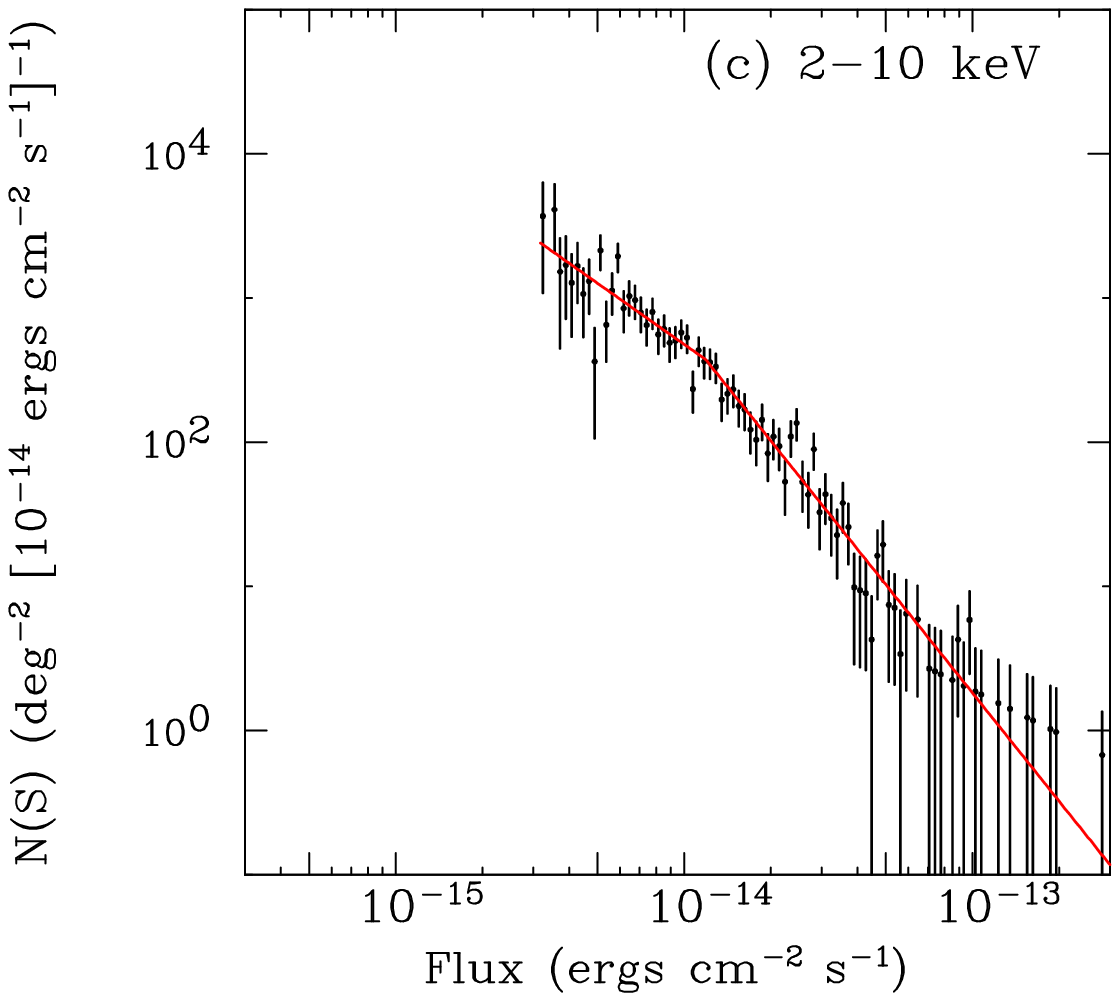}
\plotone{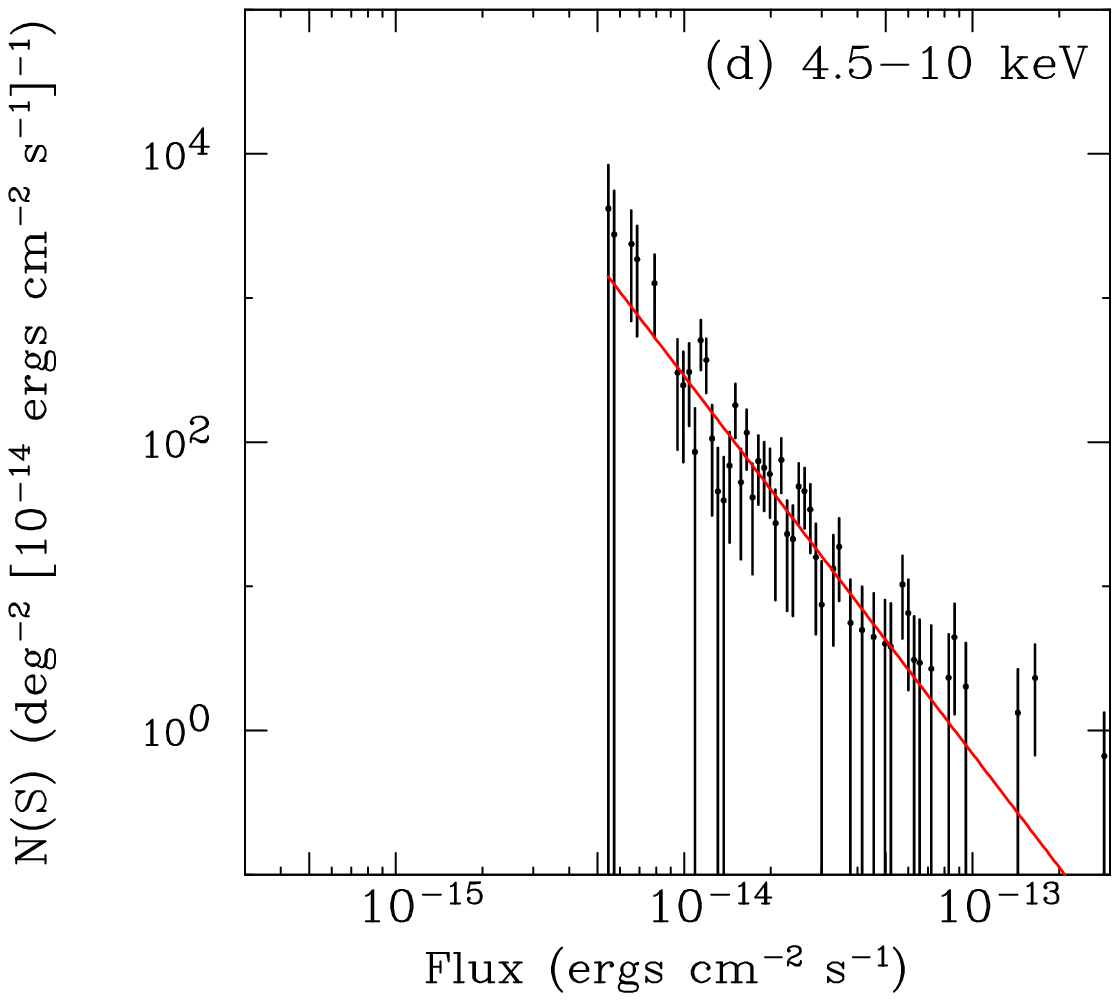}
\caption{
The differential \logn s in the (a) 0.5--2, (b) 0.5--4.5, (c)
2--10, and (d) 4.5--10 keV band. The attached errors correspond to
$1\sigma$ Poisson error one in the number of sources. The solid lines are
the best-fit broken power law model (see equation~\ref{eq:diff} and Table~7).
\label{fig9}}
\end{figure*}

The differential \logn\ are then integrated toward lower fluxes to
produce the \logn s in the integral form, $N(>S)$, where the number
density of sources with fluxes above $S$ is plotted (Figure~10). The
achieved sensitivities, defined as the fluxes at which the sky area
falls below 1\% of the maximum area (1.14 \de), are $6\times10^{-16}$,
$8\times10^{-16}$, $3\times10^{-15}$, and $5\times10^{-15}$ \ergs\ in
the 0.5--2 keV, 0.5--4.5 keV, 2--10 keV, and 4.5--10 keV band: above
these fluxes $(88\pm5)$\%, $(75\pm3)$\%, $(74\pm4)$\%, and
$(52\pm3)$\% of the XRB are resolved, when we adopt the XRB intensity
of $(7.5\pm0.4)\times10^{-12}$, $(15.3\pm0.6)\times10^{-12}$,
$(20.2\pm1.1)\times10^{-12}$, and $(12.3\pm0.7)\times10^{-12}$ \ergs\
deg$^{-2}$, respectively, taken from Table~6 of \citet{car07} with an
appropriate flux conversion for the 4.5--10 keV band. For the estimate
of the contribution of bright sources that is not well constrained
from the SXDS data, we have used the formula by \citet{car07}, who
complied the results from \chandra\ deep surveys, \xmm\ serendipitous
surveys, and the \asca\ medium sensitivity survey.

\begin{figure*}
\epsscale{0.39}
\plotone{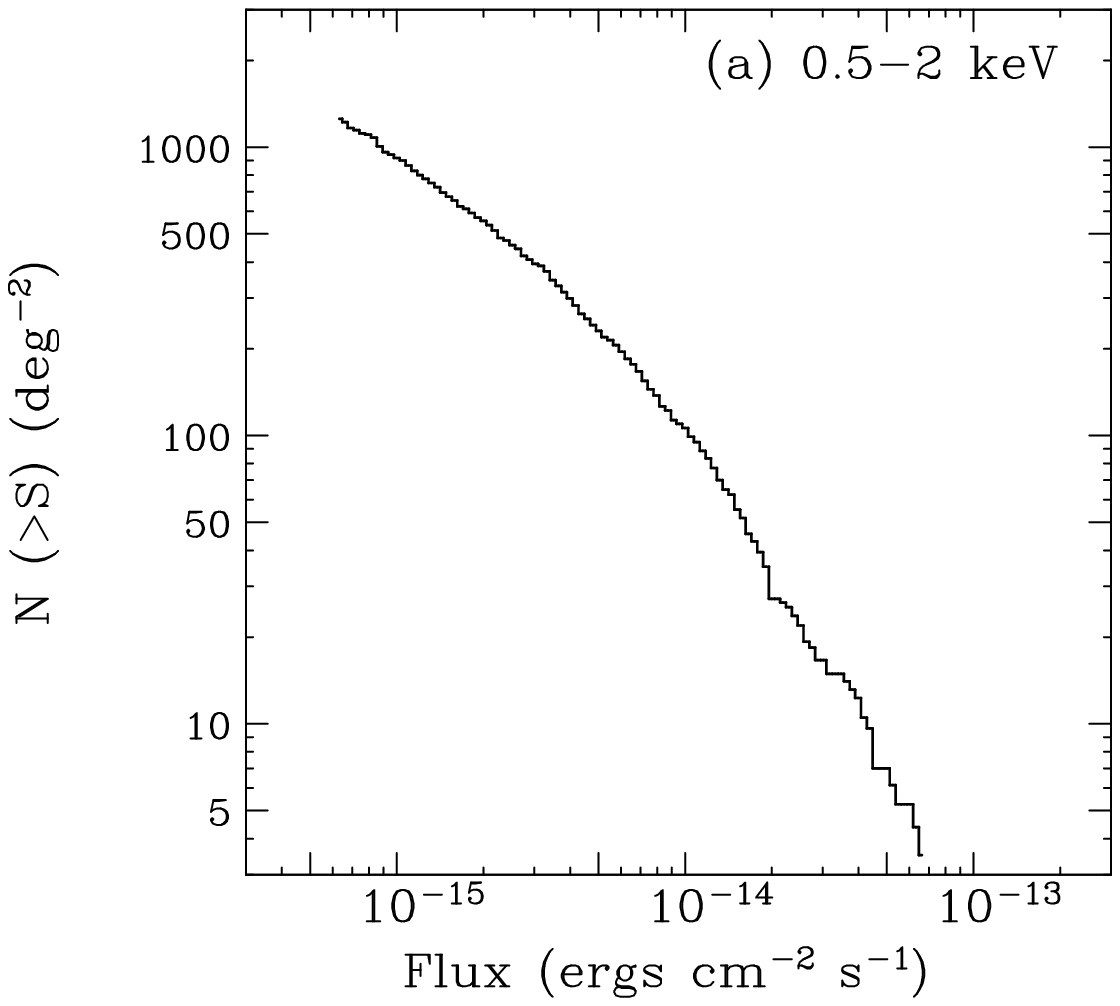}
\plotone{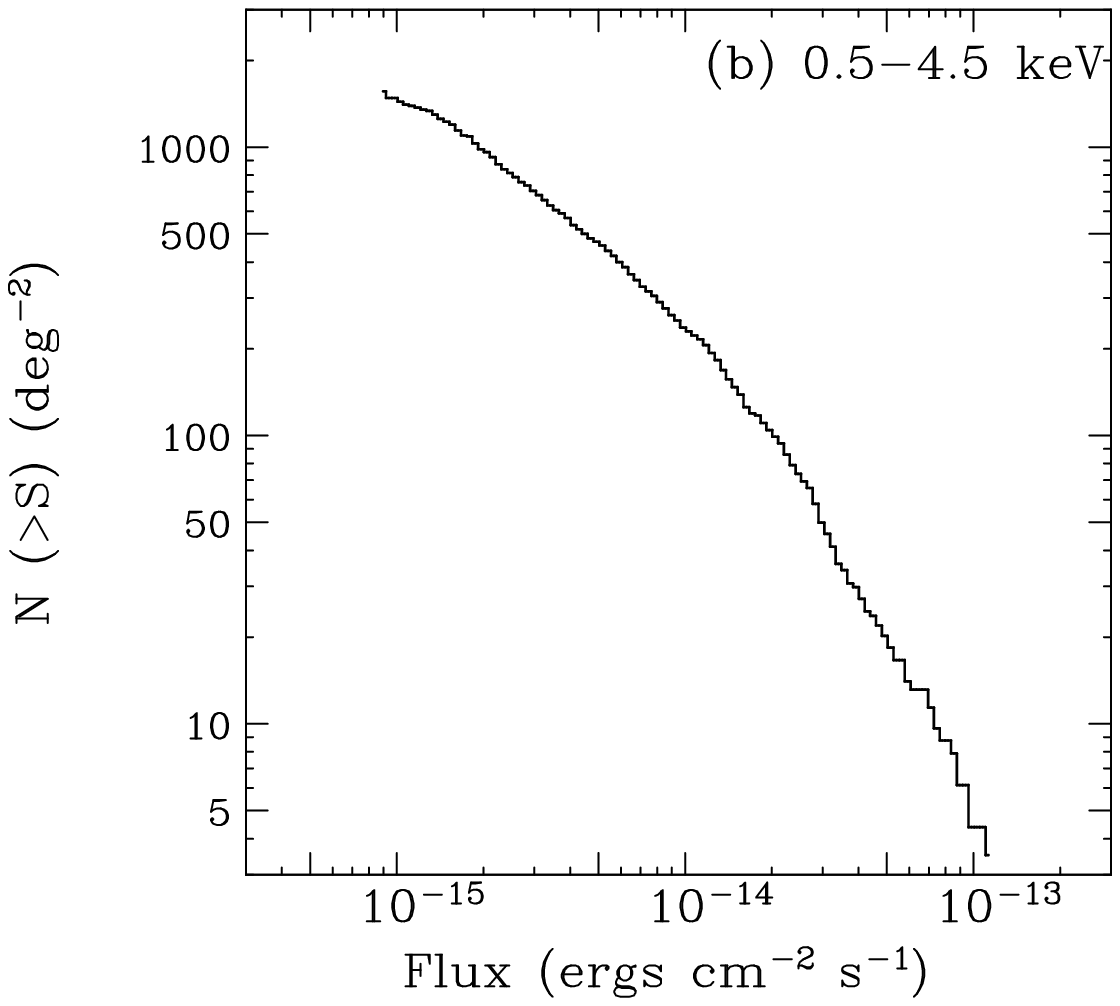}
\plotone{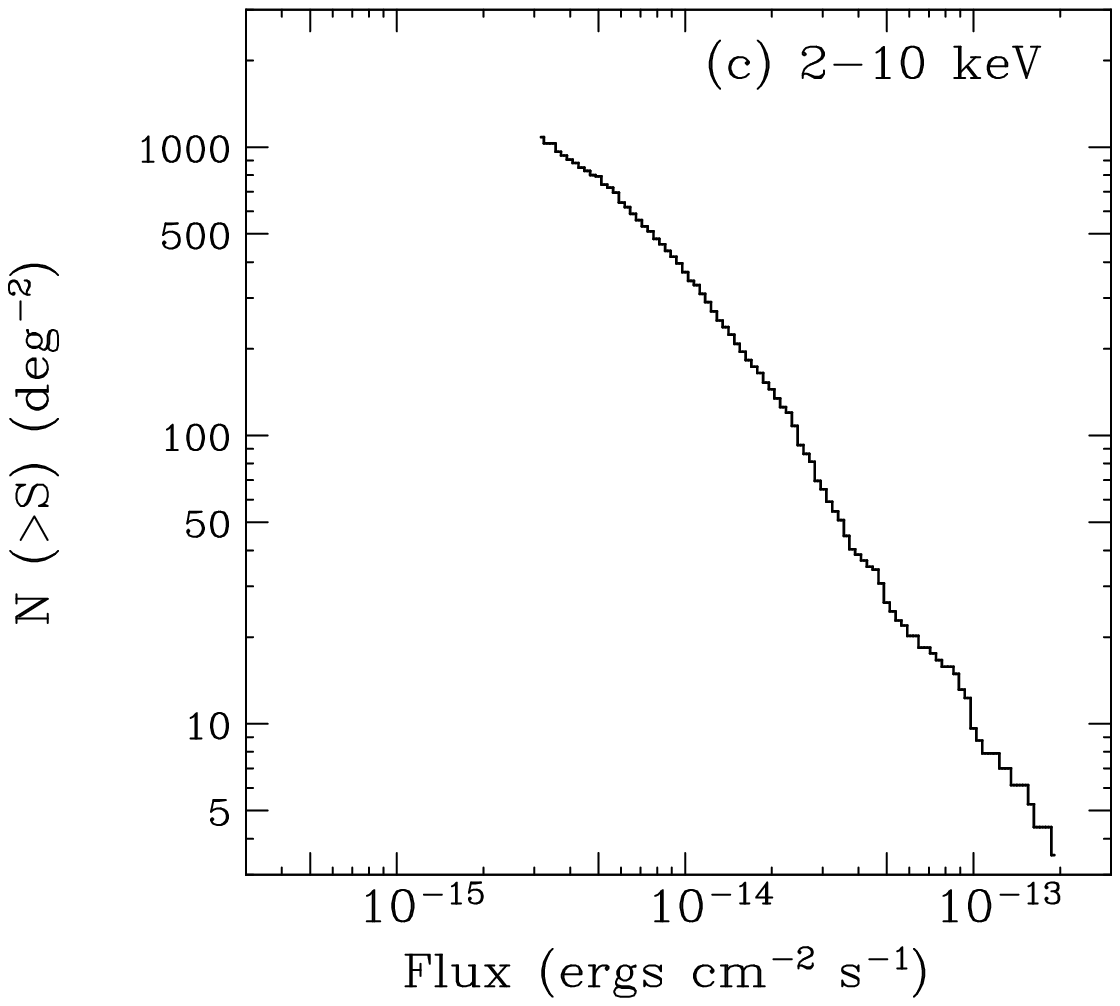}
\plotone{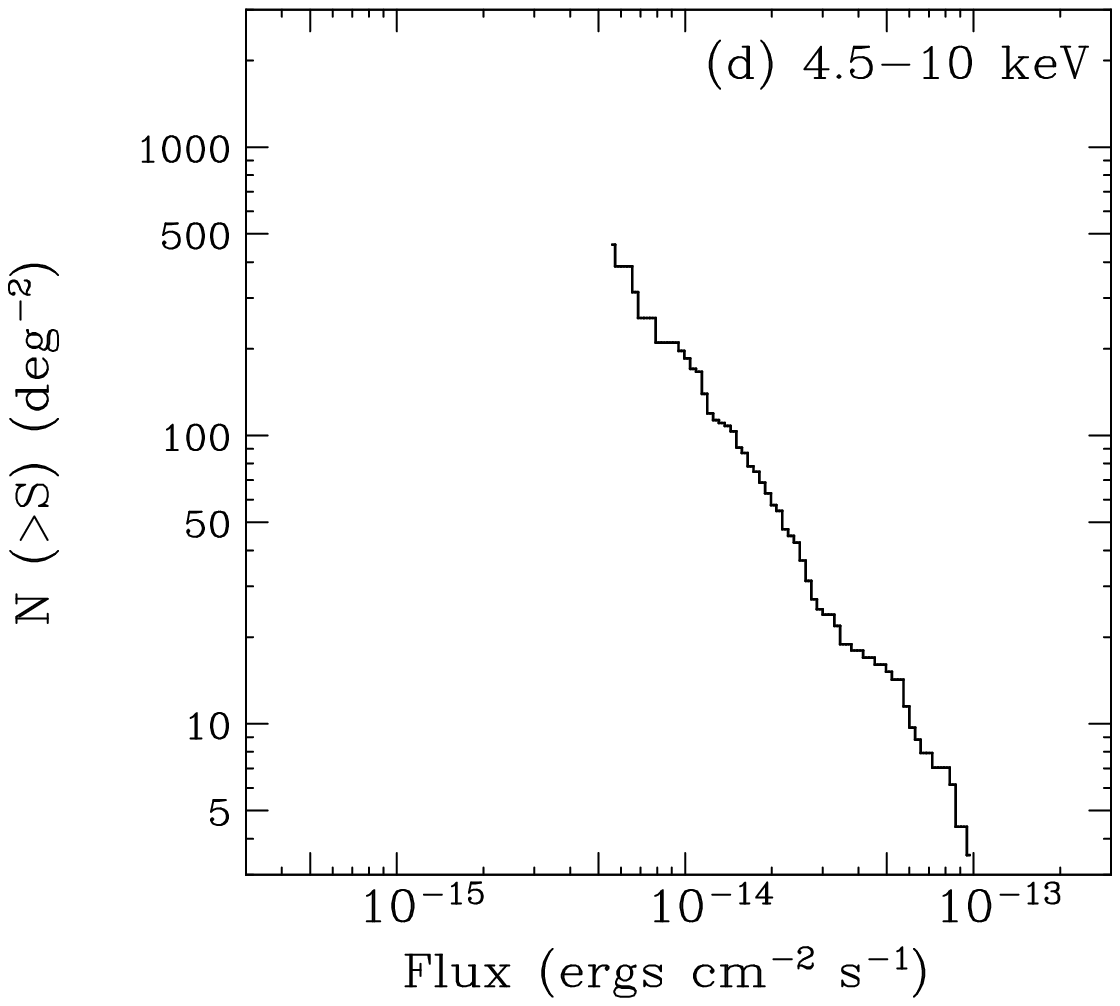}
\caption{
The \logn\ obtained from the whole SXDS field in the (a) 0.5--2, (b) 0.5--4.5, (c) 2--10, and (d) 4.5--10 keV band.
\label{fig10}}
\end{figure*}

Figure~11 shows the same \logn s scaled by $(S/10^{-14})^{1.5}$ to
stress the deviation from a Euclidean slope. The dashed curves
represent the best-fit formula by \citet{car07}; they are plotted in
the figure covering only the region where the fit is performed. Our
results are in good agreement with the previous work within
$\approx$10\%, indicating systematic errors, if any, is confined
within this level. The red points are the results of the COSMOS survey
\citep{cap07}. Here we convert the result by \citet{car07} in the
4.5--7.5 keV band and that by \citet{cap07} in the 5--10 keV band to
the 4.5--10 keV band assuming a photon index of 1.5 (the flux
conversion factor is 1.686 and 1.124, respectively). As seen from the
figure, the COSMOS source counts in the 2--10 keV are 20--30\% smaller
than ours and the \citet{car07} results. This is most probably because
while our results are truly based on the 2--10 keV survey,
\citet{cap07} make this plot from the 2--4.5 keV survey by converting
the flux into the 2--10 keV band assuming a photon index of 1.7. This
could easily miss a population of sources with hard spectra or
underestimate their fluxes. In the 0.5--2 keV band, our source counts
at $S < 3\times10^{-14}$ \ergs\ are systematically larger than both
\citet{car07} and \citet{cap07} but are consistent with the latest
results from the \xmm\ Lockman Hole survey \citep{bru08}.

\begin{figure*}
\epsscale{0.39}
\plotone{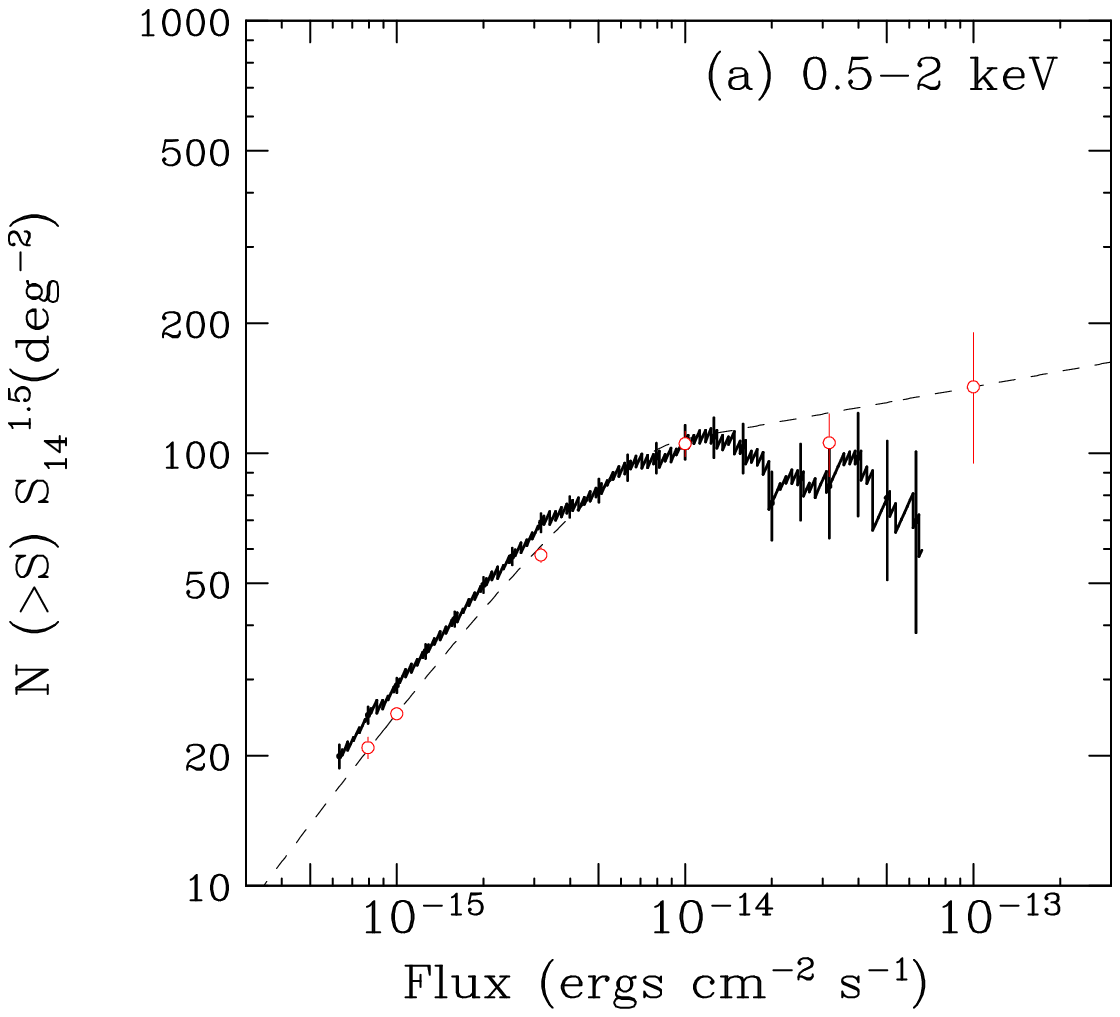}
\plotone{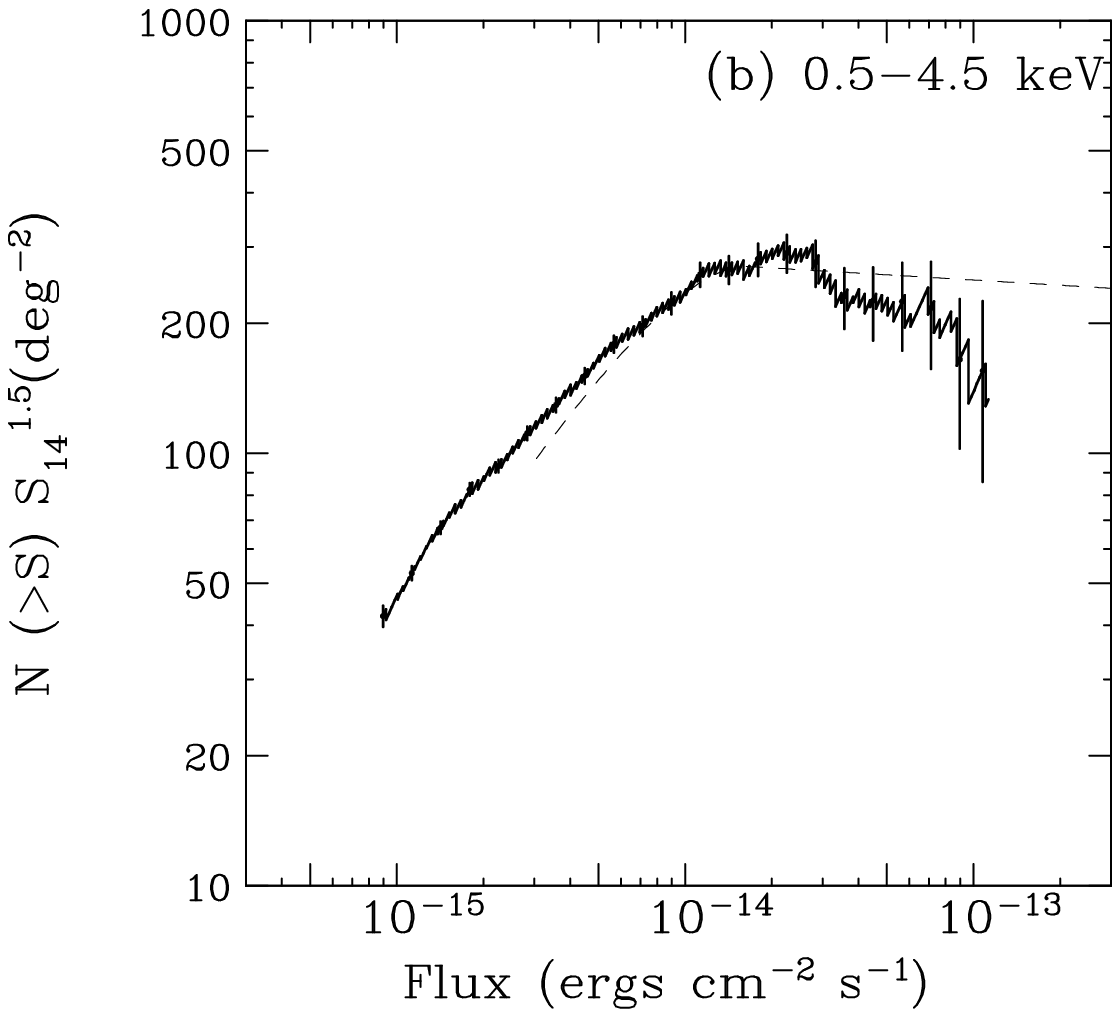}
\plotone{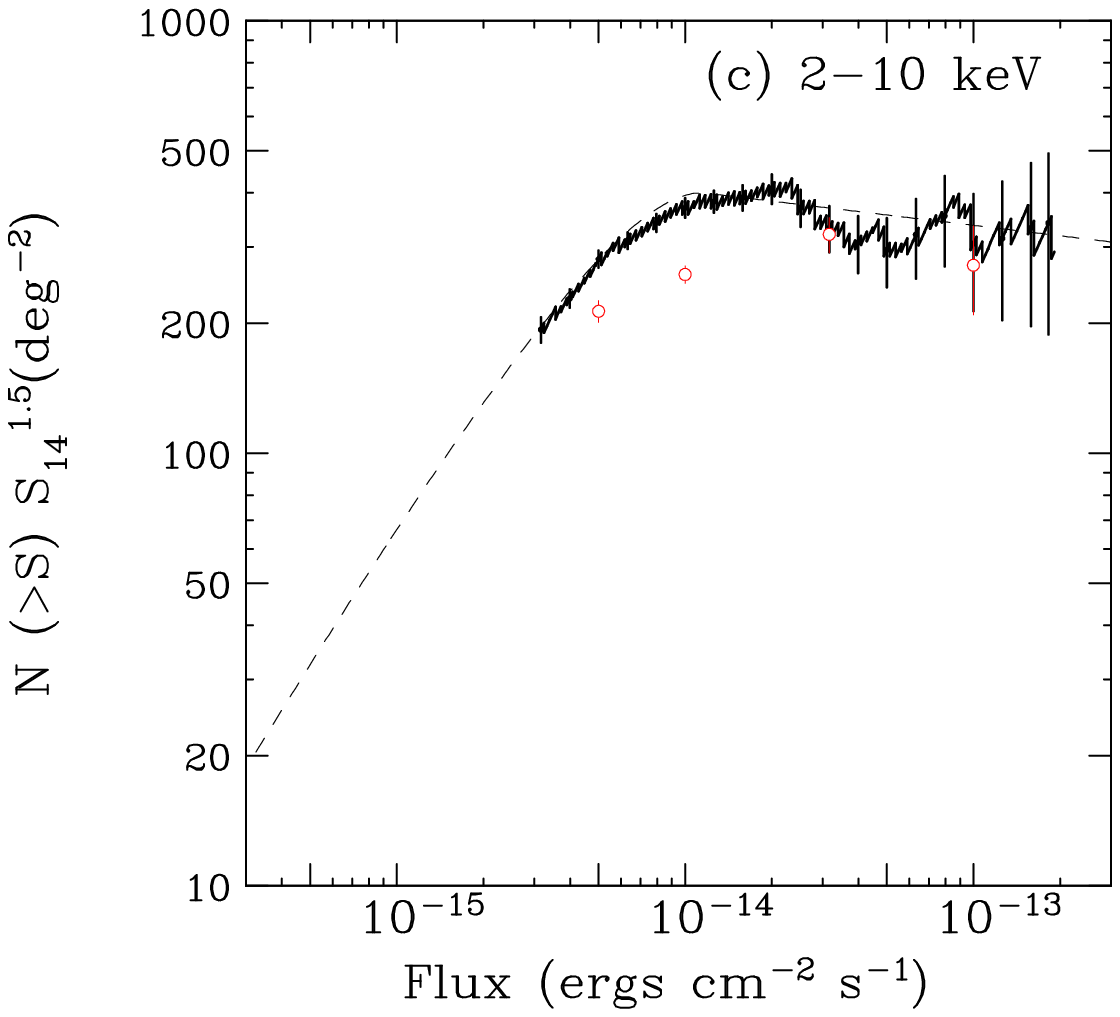}
\plotone{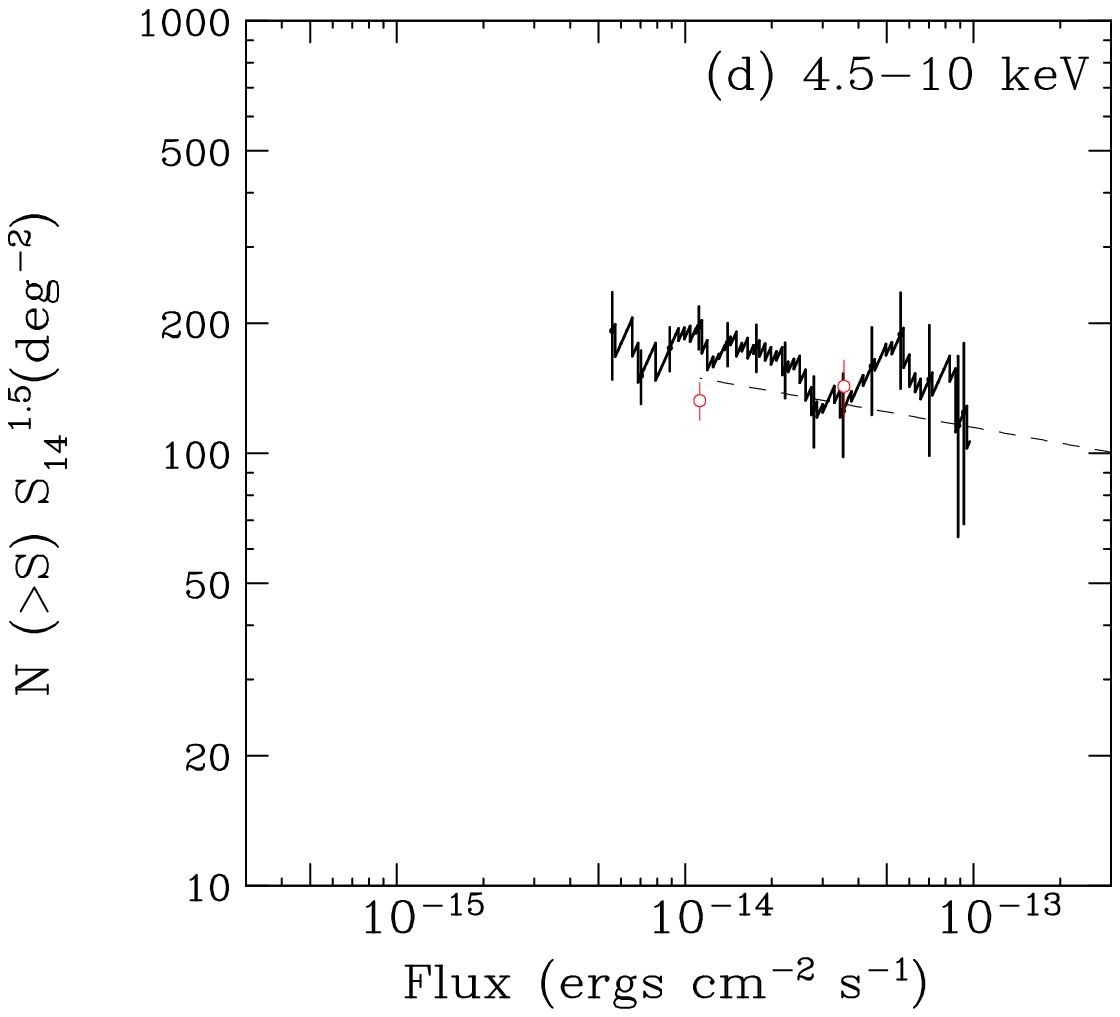}
\caption{
The \logn\ obtained from the whole SXDS field in the (a)
0.5--2, (b) 0.5--4.5, (c) 2--10, and (d) 4.5--10 keV band, scaled by
$(S/10^{-14})^{1.5}$. The error bars in our data correspond to
1$\sigma$ statistical errors. The dashed curves correspond to the
best-fit model by \citet{car07} obtained by using \chandra , \xmm ,
and \asca\ results. The open circles (red) are the results obtained from
the COSMOS survey \citep{cap07}. In (d), both results are converted
into the 4.5--10 keV band assuming a photon index of 1.5.
\label{fig11}}
\end{figure*}

\section{Clustering of Sources}

\subsection{Spatial Distribution}

Investigating the evolution of the large scale structure of the
universe is one of the main scientific objectives of the SXDS. From
X-ray surveys, we can study the clustering properties of AGNs
including both unobscured and obscured populations over a wide
redshift range as a tracer of the large scale structure. This also
helps us to understand the AGN phenomena and their environment by
measuring the mass of underlying dark matter halo where the AGN
activity took place.

Many studies have shown the presence of the significant large scale
structure in the spatial (both 2-dimensional and 3-dimensional)
distribution of X-ray sources
\citep[e.g.,][]{vik95,car98,aky00,yan03,gil03,bas04,mul04,bas05,gil05,yan06,puc06,gan06,car07,miy07}. This
leads pencil beam surveys to be inevitably subject to cosmic
variance. The SXDS, thanks to its wide and continuous area coverage,
provides us with an ideal opportunity to investigate this
issue. Figure~12 shows the location of sources detected in the 0.5--2,
0.5--4.5, or 2--10 keV band (with ML$\geq$7) in our source
catalog. Figure~13 shows the \logn\ separately derived from the seven
pointings. As done in the analysis of \S~3.3, here we excluded any data
with shorter exposures in overlapping regions of multiple pointings,
and hence the results are statistically independent one another. It
can be seen that the source counts show significant variation among
the pointings, depending on the flux limit. In particular, those of
SDS7 in the 0.5--2 and 0.5--4.5 keV band are by more than 30\% smaller
than the other fields in the medium to high flux range. These results
indicate that the cosmic variance indeed exists over an area scale of
one \xmm\ FOV, $\sim$0.2 deg$^2$, which is much larger than the FOV of
\chandra\ observatory.

\begin{figure}
\epsscale{1.0}
\plotone{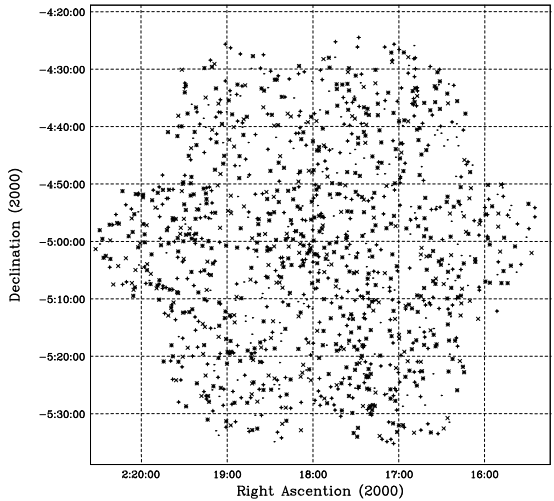}
\caption{
The spatial distribution of detected X-ray sources in the whole
 SXDS field. Crosses, dots, and diagonal crosses correspond to 
those detected in the 0.5--2 keV, 0.5--4.5 keV, and 2--10 keV band, 
respectively.
\label{fig12}}
\end{figure}

\begin{figure*}
\epsscale{0.39}
\plotone{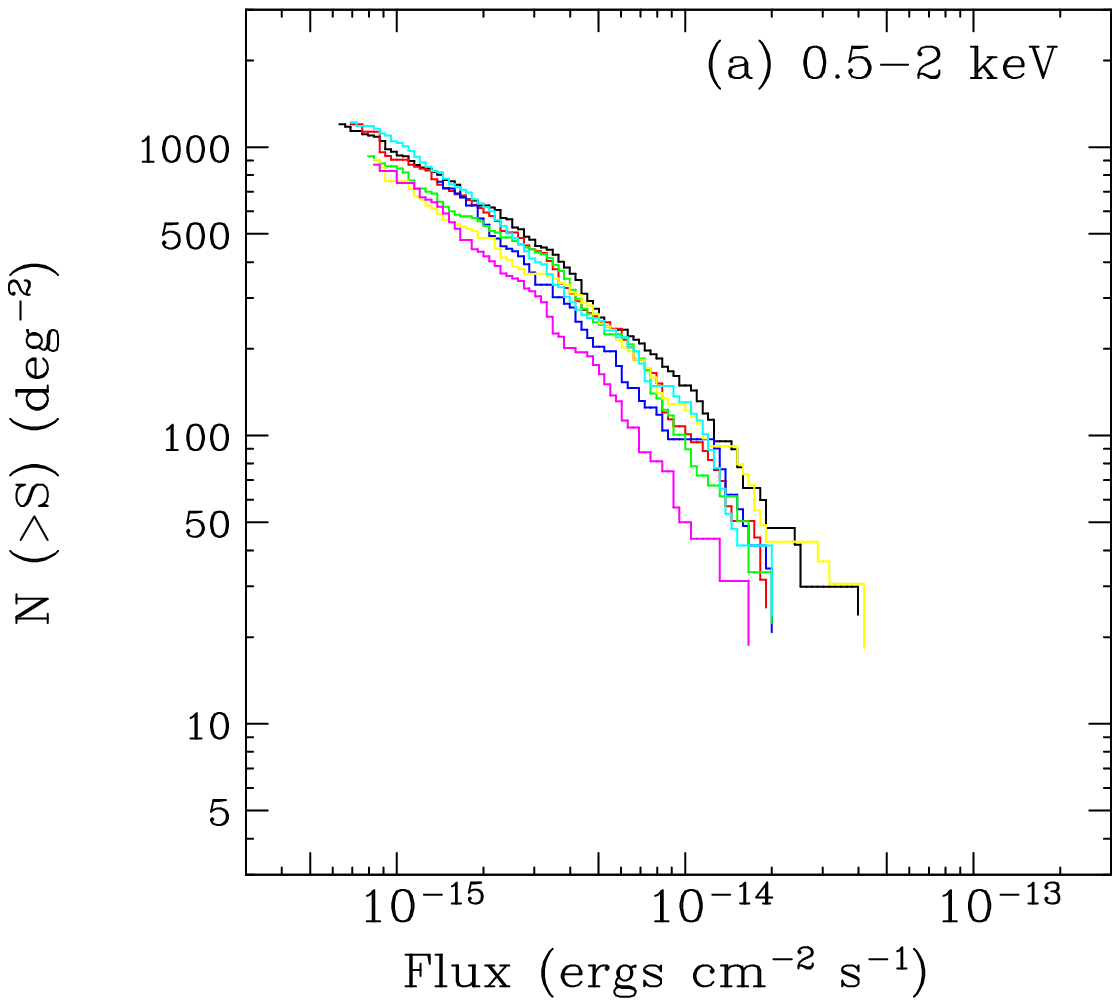}
\plotone{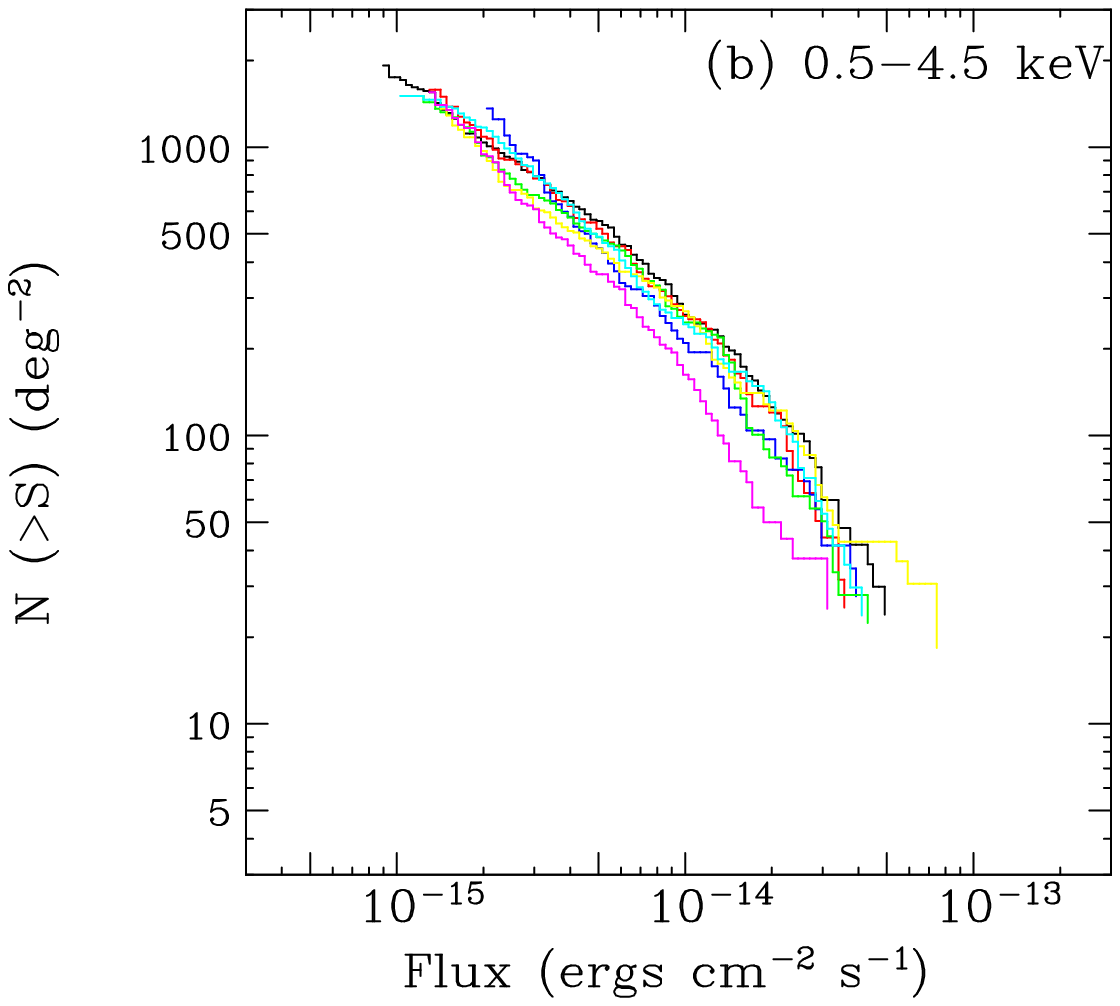}
\plotone{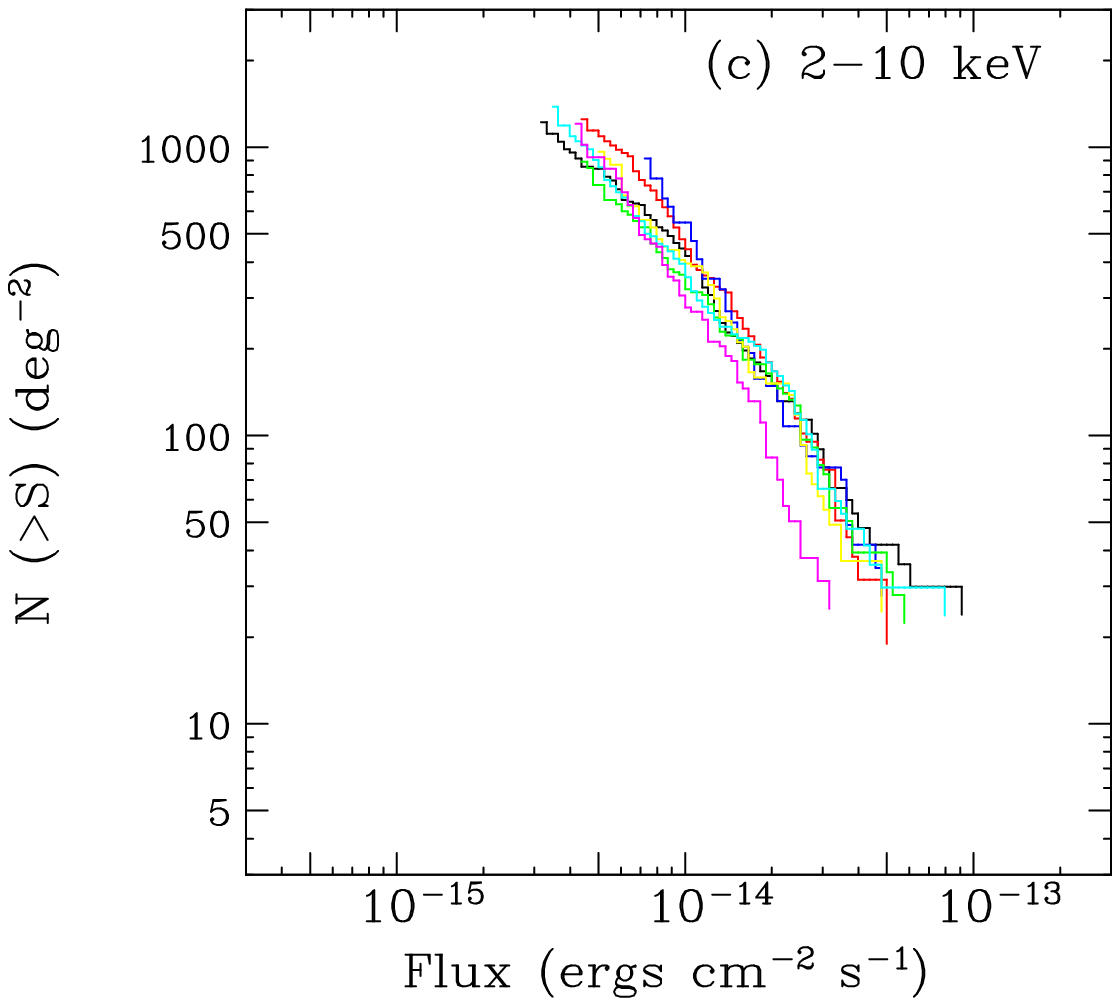}
\plotone{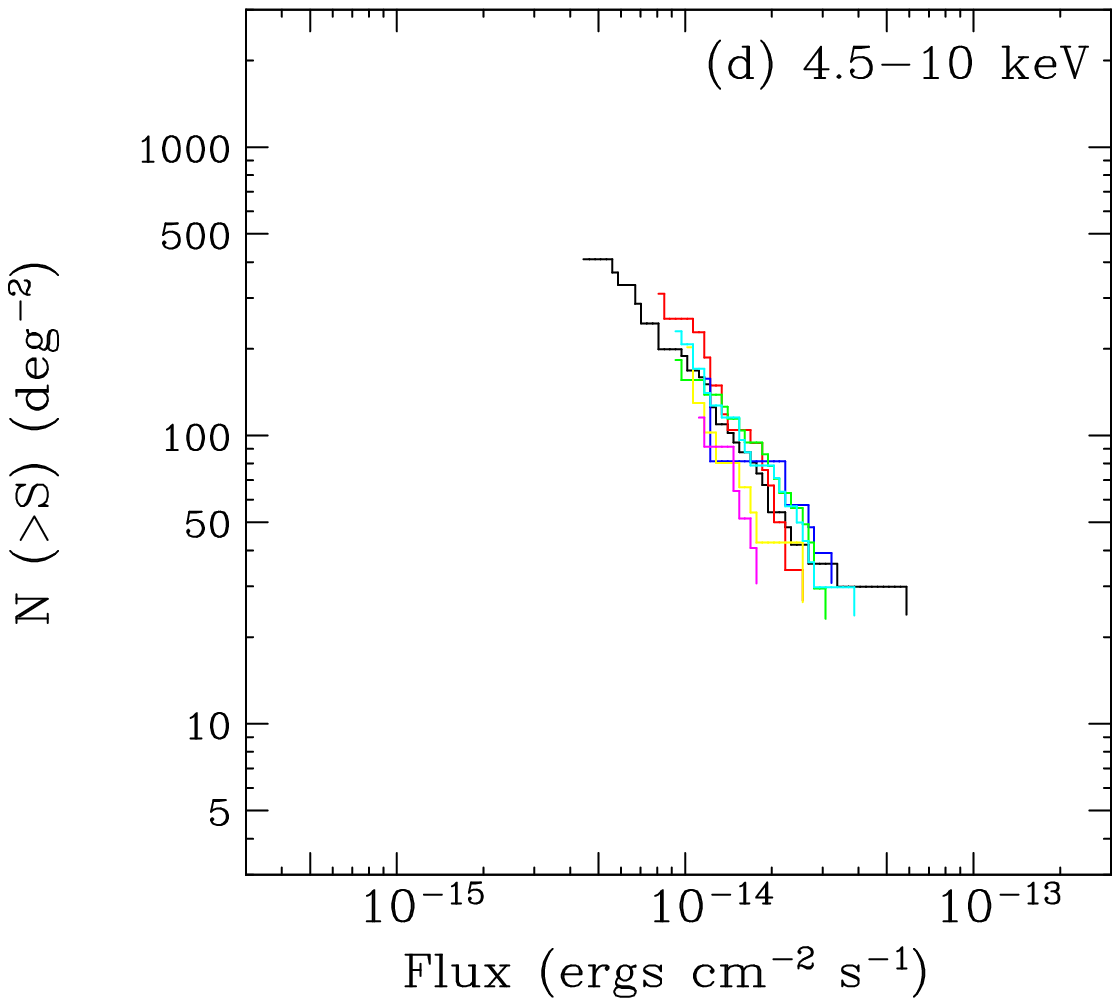}
\caption{
The \logn\ for each field in the (a) 0.5--2, (b) 0.5--4.5, (c) 
2--10, and (d) 4.5--10 keV band. Black: SDS-1, Red: SDS-2,
 Blue: SDS-3, Green: SDS-4, Yellow: SDS-5, Cyan: SDS-6, Magenta: SDS-7.
In the overlapping regions between different pointings data of 
lower exposure are excluded in the analysis i.e., the plots are 
statistically independent of one another.
\label{fig13}}
\end{figure*}

\subsection{Auto Correlation Function}\label{sec:ACF}

The most straightforward approach to quantify the clustering properties
of sources is to calculate the angular two-point correlation function (or auto-correlation function; hereafter ACF). The ACF, $w(\theta)$, is defined as
$$
dP = n^2[1+w(\theta)]d\Omega_1 d\Omega_2,
$$
where dP gives the probability of finding a pair of two objects
located at the differential solid angle of $\Omega_1$ and $\Omega_2$
with a separation angle of $\theta$, and $n$ is the mean number
density. 

To estimate the ACF, we essentially follow the same procedure as done
in \citet{miy07}. We adopt the \citet{lan93} estimator,
$$
w_{\rm obs}(\theta) = \frac{DD-2DR+RR}{RR}, 
$$
where $DD$, $DR$, $RR$ are the normalized number of data-data pairs,
data-random pairs, and random-random pairs, respectively. In
calculating the ACF, it is critical to ensure that the sensitivity map
is correctly modeled at every position. Hence, considering possible
position-dependent systematic errors in the sensitivity maps created
in \S~4.1, we introduce a ``safety factor'', $f=1.1$, to multiply the
nominal sensitivity limits and select only sources whose observed flux
is higher than the (corrected) sensitivity limit at the detected
position. The choice of $f=1.1$ is quite conservative, and
indeed we obtain essentially the same results even when $f=1.0$ is
adopted. We make a random sample whose number is 10 times that of the
actual data. The flux distribution of the random sample is taken to be
the same as in the actual source list, and their positions are
randomly distributed by satisfying that the flux must be higher than
the sensitivity limit at the allocated position.

Figure~14 show our results of the ACF obtained in the 0.5--2 and 2--10
keV bands. The flux range and number of sources used in the analysis
are summarized in the 2nd and 3rd columns of Table~8, respectively.
The attached error is estimated by simulation: we produce 100 sets of
random samples with the same source number as in the data, obtain a
standard deviation of the ACF signal obtained by the same procedure as
applied for the actual data, and then multiplied it by
$\sqrt{1+w(\theta)}$. We fit the ACF with a power law form
$$
w (\theta) = (\theta/\theta_c)^{1-\gamma}
$$ 
in the range of $\theta=0.5-10$ \hun , considering the size of the PSF
of \xmm . We fix $\gamma=1.8$ \citep[][]{pee80}, which is difficult
to constrain by our data, and derive the correlation length $\theta_c$
as a free parameter. To take into account the coupling of the ACF
between different bins, we utilize the ``covariance matrix'', which is
also obtained from the simulation of the random sample described
above; we refer the readers to \S~3.4 of \citet{miy07} for details. In
the fit, we also consider an offset produced by the integral
constraint, $\int \int w(\theta) d\Omega_1 d\Omega_2 = 0 $, assuming
that the power law form of the ACF holds over the whole area. The
obtained values of $\theta_c$ are listed in the 6th column of Table~8.

\begin{deluxetable*}{ccccccccc}
\tabletypesize{\footnotesize}
\tablenum{8}
\tablecaption{Clustering Properties of the X-ray Sources\label{tbl-8}} 
\tablehead{
\multicolumn{3}{c}{Sample}&\colhead{}&
\multicolumn{5}{c}{Clustering Properties}\\
\cline{1-3} \cline{5-9} \\
\colhead{Detection}& \colhead{Flux Range} &\colhead{Number of}&\colhead{}&
\colhead{$\theta_{\rm min}$-$\theta_{\rm max}$}&\colhead{$\gamma$}&
\colhead{$\theta_{\rm c}$\tablenotemark{a,d}}&\colhead{$z_{\rm eff}$\tablenotemark{b}}&\colhead{$r_c$\tablenotemark{c,d}}\\
\colhead{Band}& \colhead{($10^{-14}$ erg s$^{-1}$ cm$^{-2}$)}&\colhead{Sources}&\colhead{}&
\colhead{(\hun)}&
\colhead{(fixed)}&\colhead{(\byo)}&\colhead{}&\colhead{($h^{-1}$ Mpc)}}
\startdata
0.5--2 keV& 0.06--22& 765 &&
0.5--10 & 
1.8& 5.9$^{+1.0}_{-0.9}$ & 1.3 & 14.9$\pm$1.1 \\
2--10 keV & 0.32--33& 573 && 
0.5--10 & 
1.8& 0.1 ($<1.5$) & 1.1 & 2.3 ($<7.6$)
\enddata
\tablenotetext{a}{The 2-dimensional correlation length, assuming a power law
form of $(\frac{\theta}{\theta_c})^{1-\gamma}$ 
for the angular correlation function. The fit is performed between $\theta_{\rm min}$ and $\theta_{\rm max}$ with the integral constraint (see text).}
\tablenotetext{b}{The median redshift contributing to the angular correlation.}
\tablenotetext{c}{The correlation length assuming the 
comoving model ($\epsilon=\gamma-3$), converted from $\theta_{\rm c}$ 
via the Limber transformation (see text).}
\tablenotetext{d}{The error is $1\sigma$ while the upper limit is 90\% confidence limit.}
\end{deluxetable*}

\begin{figure}
\epsscale{1.0}
\plotone{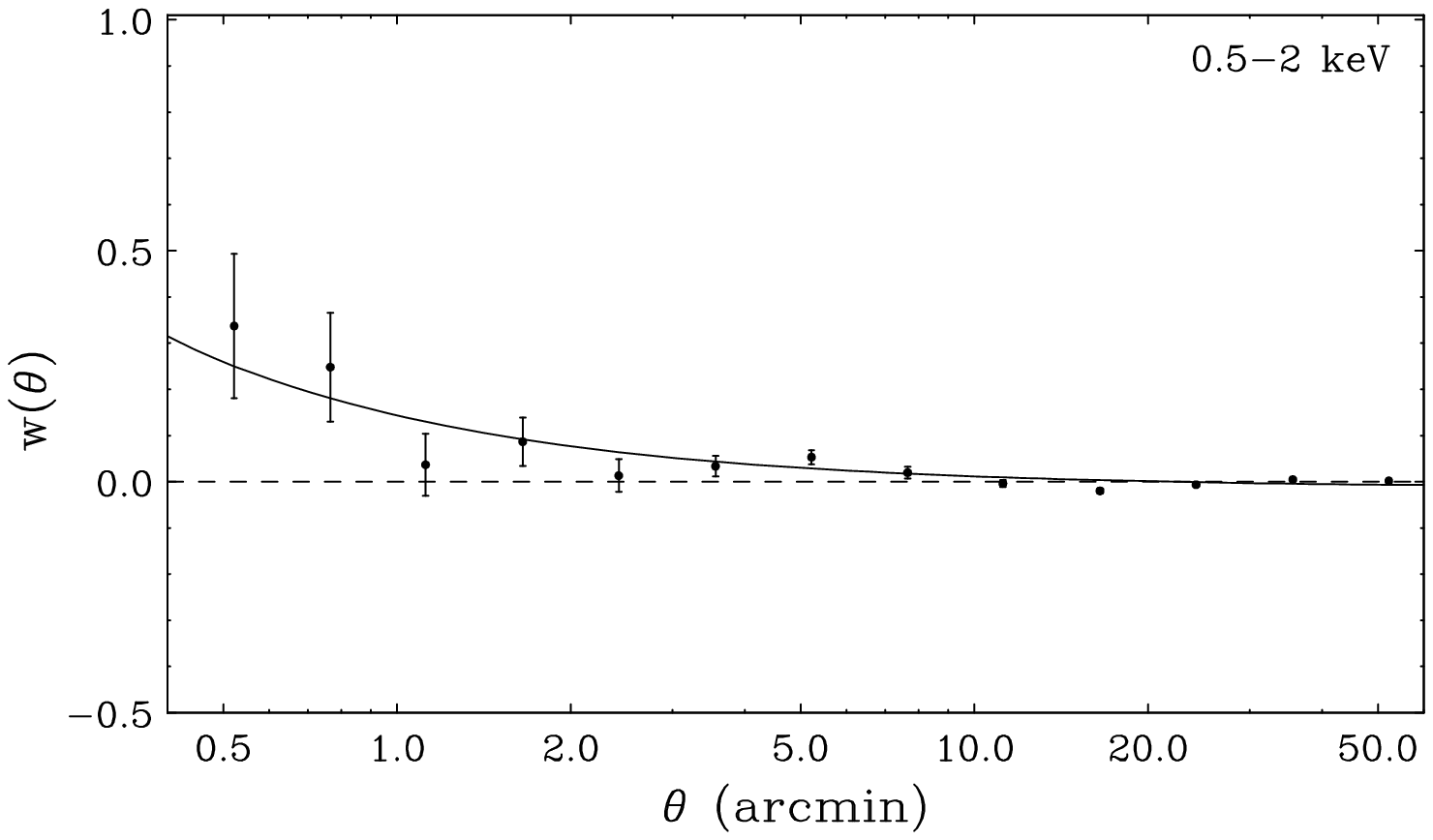}
\plotone{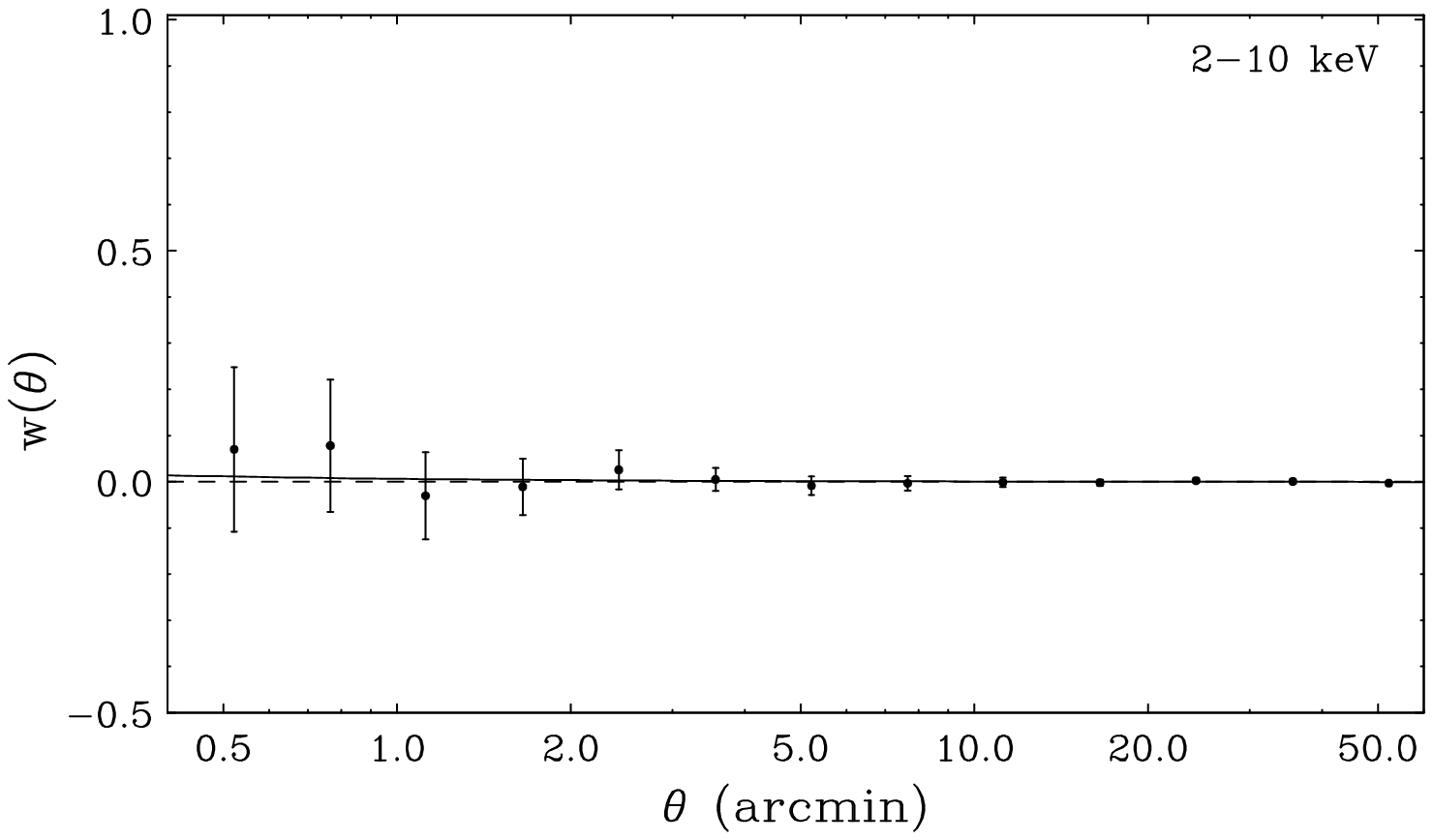}
\caption{
The auto-correlation function of the SXDS sources in the 
0.5--2 keV and 2--10 keV bands. The errors are 1
 $\sigma$. The line represents the best-fit model (power law plus constant
to account for the integration constraint).
\label{fig14}}
\end{figure}

We detect an ACF signal from the 0.5--2 keV selected sample with
$\theta_c=5.9$\byo $^{+1.0^{{\prime}{\prime}}}_{-0.9^{{\prime}{\prime}}}$. The signal
is not significantly found from the 2--10 keV sample, however, with a
90\% confidence upper limit of $\theta_c<$ 1.5\byo. By Limber's equation
\citep{pee80}, the ACF can be transformed to the 3-dimensional
correlation function
$$
\xi(r,z) = (r/r_{c,0})^{-\gamma} (1+z)^{-3-\epsilon+\gamma},
$$ once the redshift distribution of the sample is known. 
Here we assume
the comoving clustering model where the correlation length $r_c,0$ is
constant in the comoving coordinates, hence $\epsilon = \gamma - 3$.
Figure~15 shows the AGN redshift distribution for
the 0.5--2 and 2--10 keV samples estimated from the model by
\citet{ued03}. From the $\theta_c$ values and redshift distribution,
we finally obtain $r_{c,0}=14.9\pm1.1$ $h^{-1}$ Mpc and
$r_{c,0}<7.6$ $h^{-1}$ Mpc with a medium redshift of 1.3 and 1.1
for the 0.5--2 and 2--10 keV selected sample, respectively.  The bias
parameter, the ratio of the rms fluctuation amplitude $\sigma_8$ 
\citep{pee80} between AGNs and underlying mass, is found to be about
5.8$\pm0.5$ for the 0.5--2 keV selected AGNs at a median redshift of
1.3.

\begin{figure}
\epsscale{1.0}
\plotone{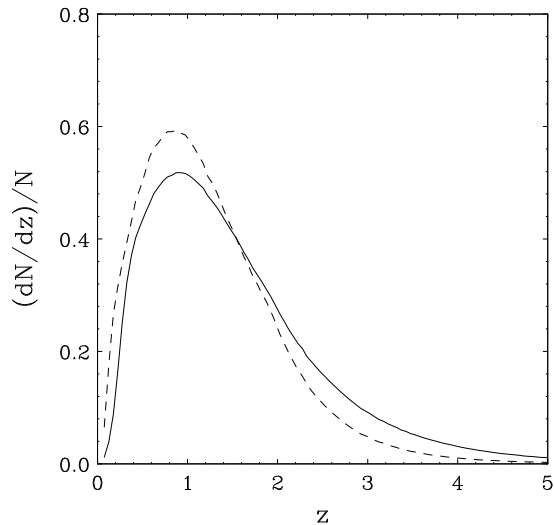}
\caption{
The normalized redshift distribution of AGN in our samples selected in
the 0.5--2 keV (solid curve) and 2--10 keV (dashed curve) bands,
estimated from the model by \citet{ued03}.
\label{fig15}}
\end{figure}

The correlation length in the 0.5--2 keV is consistent with the previous
results from soft X-ray surveys at similar redshift range reported by
e.g., \citet{bas05}, \citet{puc06}, and \citet{car07}, although it is
larger than those obtained from the optically selected QSOs
\citep[][]{cro05}. By contrast to several previous works that reported
strong correlation of hard-band ($>$2 keV) selected samples
\citep{yan03,bas04}, we find no evidence for significant clustering
signals in the hard band. The detection of ACF in the 0.5--2 keV band
but not in the 2--10 keV band is similar to what was found by
\citet{gan06} and \citet{car07}. It is curious that the first result
from the XMM-COSMOS reports a relatively large correlation length for
the 4.5--10 keV band sample but smaller for the 2--4.5 keV band sample
\citep{miy07}. Our 2--10 keV result is consistent with the COSMOS 2--4.5
keV result. We do not find significant ACF signals from the 4.5--10 keV
sample, although the result is more subject to the statistical
fluctuation due to the small number of sources detectable in this energy
band.

\section{Long Term X-ray Variability of Sources}

We investigate the long term variability of X-ray sources detected in
the SDS-4 field, for which two observations were performed with a time
separation of about 2.5 year. The exposures of both observations are
similar and are sufficiently long ($>20$ ksec). Using the X-ray catalog
produced from the combined images of the two observations, we examine
the fluxes of the cataloged sources separately in each epoch by fixing
their positions. To ensure high signal-to-noise ratio, we only use
sources whose summed ML value from the 0.5--2 keV and 2--4.5 keV bands
exceeds 15 in the combined catalog and vignetting-corrected exposure is
longer than 15 ksec in both observations. Figure~16 shows the comparison
of the 0.5--4.5 keV flux between the first and second epochs for this
sample. As noticed from the figure, many sources at intermediate fluxes
that have sufficiently small error bars show a significant variability,
demonstrating the impact of time variability in studying AGN
properties. The results are consistent with previous studies
\citep{pao04,mat07} reporting that the fraction of variable sources on
time scale of months to years is 80--90\% or higher in \chandra\ and
\xmm\ deep fields.

\begin{figure}
\epsscale{1.0}
\includegraphics[angle=270,scale=0.4]{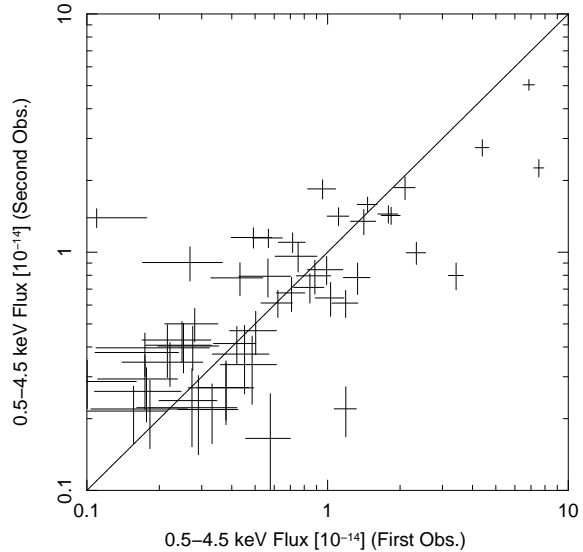}
\caption{
The comparison of the 0.5--4.5 keV flux measured in the two different epochs
for sources detected in the SDS-4 field.
\label{fig16}}
\end{figure}

\section{Summary}

We have presented the first X-ray source catalog and their basic X-ray
properties in the SXDS field, based on the seven \xmm\ pointings
performed between 2000 and 2003 that cover a continuous area of 1.14
\de\ in the 0.2--10 keV band. The catalog consists of 1245 sources in
total, consisting of those detected either in the 0.3--0.5 keV, 0.5--2
keV, 2--4.5 keV, 4.5--10 keV, 0.5--4.5 keV, and 2--10 keV with
detection likelihood larger than 7. From empirical sensitivity maps
derived by simplified simulation, we obtain \logn s in various energy
bands bands, which bridge the flux range between \chandra\ deep
surveys and brighter surveys. Clustering properties of X-ray sources
are also studied by means of auto correlation function. We 
detect significant signals that can be fit by
$(\theta/\theta_c)^{-0.8}$ with $\theta_c=5.9$\byo $^{+1.0^{{\prime}{\prime}}}_{-0.9^{{\prime}{\prime}}}$
for the 0.5--2 keV selected sample at an estimated median redshift of
$z_{\rm eff}\approx 1.3$, and an 90\% upper limit of $\theta_c <
1.5$\byo\ for the 2--10 keV selected sample at $z_{\rm eff}\approx
1.1$. Two pointing data separated by 2.5 years indicate the importance
of flux variability on a time scale of years. Our results establish
the average properties of X-ray sources at fluxes where a majority of
the XRB is produced, being least affected by cosmic variance. The data
presented here constitute a major component of the SXDS project and
shall be used for a number of research programs in combination with
other wavelengths data.

\acknowledgments

This work was done as a part of the large international collaboration
of the SXDS project. We thank the members of the \xmm\ SSC and the
\subaru\ observatory project team. We also thank Dr. Takamitsu Miyaji
for his invaluable advise in the calculation of the auto correlation
function. Part of this work was financially supported by Grants-in-Aid
for Scientific Research 17740121, and by the Grant-in-Aid for the 21st
Century COE ``Center for Diversity and Universality in Physics'' from
the Ministry of Education, Culture, Sports, Science and Technology
(MEXT) of Japan.

\vspace{3cm}
\begin{center}
\begin{LARGE}
Full machine-readable ascii file of Table~2 is available at

http://www.kusastro.kyoto-u.ac.jp/\~{}yueda/sxds/tab2
\end{LARGE}
\end{center}

\clearpage
\LongTables
\begin{landscape}
\tablefontsize{\tiny}
\begin{deluxetable}{lrrcccccccccccccc ccccc}
\tabletypesize{\scriptsize}
\tablewidth{0pt}
\setlength{\tabcolsep}{0.01in}
\tablenum{2}
\tablecaption{The X-ray source list of the SXDS\label{tbl-2}}
\tablehead{
\colhead{}&
\multicolumn{3}{c}{Position (J2000)}&
\multicolumn{6}{c}{Likelihood}&
\multicolumn{4}{c}{Count Rate}&
\multicolumn{3}{c}{Hardness Ratio}&
\colhead{}&\colhead{}&\colhead{}&\colhead{}&\colhead{}\\
\colhead{\#}&
\colhead{R.A.}& \colhead{Dec.}& \colhead{err}&
\colhead{0.3-0.5}& \colhead{0.5-2}& \colhead{2-4.5}& \colhead{4.5-10}& \colhead{0.5-4.5}& \colhead{2-10}&
\colhead{0.3-0.5}& \colhead{0.5-2}& \colhead{2-4.5}& \colhead{4.5-10}& 
\colhead{$HR1$}& \colhead{$HR2$}& \colhead{$HR3$}&
\colhead{Field}&\colhead{Offset}&\colhead{Exp.}&\colhead{Bgd.}&\colhead{Note}
\\
\colhead{XMM}&
\colhead{(deg)}& \colhead{(deg)}& \colhead{($''$)}&
\multicolumn{6}{c}{}&
\multicolumn{4}{c}{(c ksec$^{-1}$)}&
\multicolumn{3}{c}{}&
\colhead{(a)}& \colhead{(b)}& \colhead{(c)}& \colhead{(d)}& \colhead{(e)}\\
}
\startdata
0001 & 33.85387& $-$4.92870&2.36&    0.0& 19.9&  3.2&  0.4& 26.5&  5.1&  0.00$\pm$ 0.09& 3.45$\pm$ 0.79& 1.39$\pm$ 0.63& 0.86$\pm$ 0.81& 1.00$\pm$0.05&$-$0.42$\pm$0.21&$-$0.24$\pm$0.49 &5 &15.6 & 9.6 &2.6 &\\
0002 & 33.85392& $-$4.90221&2.24&   16.1& 33.7&  2.5&  0.8& 37.1&  6.2&  2.69$\pm$ 0.74& 6.33$\pm$ 1.25& 1.38$\pm$ 0.72& 1.18$\pm$ 0.90& 0.40$\pm$0.14&$-$0.64$\pm$0.16&$-$0.08$\pm$0.46 &5 &16.0 & 5.5 &2.6 &\\
0003 & 33.86346& $-$5.03354&2.88&    0.3&  0.0&  0.0&  6.8&  0.4&  7.5&  0.24$\pm$ 0.27& 0.18$\pm$ 0.31& 0.09$\pm$ 0.25& 2.72$\pm$ 0.82& $-$0.13$\pm$1.01&$-$0.33$\pm$1.43&0.94$\pm$0.17 &5 &14.8 &12.2 &2.6 &\\
0004 & 33.87124& $-$4.96292&2.99&    5.2& 14.1&  0.0&  0.0& 11.1&  0.0&  0.79$\pm$ 0.34& 2.33$\pm$ 0.57& 0.00$\pm$ 0.11& 0.00$\pm$ 0.28& 0.49$\pm$0.19&$-$1.00$\pm$0.09&\nodata &5 &14.2 &15.5 &2.6 &\\
0005 & 33.87257& $-$4.89099&1.85&    2.5& 11.8&  6.7&  0.7& 22.4&  8.6&  0.68$\pm$ 0.38& 2.85$\pm$ 0.73& 1.76$\pm$ 0.55& 0.88$\pm$ 0.69& 0.61$\pm$0.19&$-$0.24$\pm$0.19&$-$0.33$\pm$0.38 &5 &15.2 & 9.7 &2.6 &\\
0006 & 33.87878& $-$4.98760&2.84&    0.0&  2.6&  2.9&  0.0&  8.7&  2.8&  0.00$\pm$ 0.04& 0.89$\pm$ 0.41& 0.61$\pm$ 0.29& 0.00$\pm$ 0.38& 1.00$\pm$0.10&$-$0.19$\pm$0.32&$-$1.00$\pm$1.24 &5 &13.6 &17.9 &2.5 &\\
0007 & 33.89418& $-$5.06675&0.56&  131.3&554.7&  5.5&  0.0&484.2&  4.3&  6.86$\pm$ 0.88&25.08$\pm$ 1.56& 1.20$\pm$ 0.43& 0.03$\pm$ 0.25& 0.57$\pm$0.05&$-$0.91$\pm$0.03&$-$0.94$\pm$0.40 &5 &13.6 &14.1 &2.8 &\\
0008 & 33.90476& $-$4.94904&1.25&   24.9& 41.6&  6.7&  0.0& 50.9&  5.9&  1.56$\pm$ 0.33& 3.57$\pm$ 0.56& 0.85$\pm$ 0.27& 0.00$\pm$ 0.21& 0.39$\pm$0.11&$-$0.62$\pm$0.11&$-$0.99$\pm$0.49 &5 &12.3 &16.8 &2.2 &\\
0009 & 33.90754& $-$5.08932&2.64&    0.0&  6.8&  3.4&  0.0& 13.6&  3.0&  0.00$\pm$ 0.06& 1.75$\pm$ 0.56& 0.77$\pm$ 0.34& 0.14$\pm$ 0.32& 1.00$\pm$0.06&$-$0.39$\pm$0.23&$-$0.69$\pm$0.61 &5 &13.4 &17.3 &2.7 &\\
0010 & 33.91171& $-$5.01450&1.30&   10.6& 72.1&  6.0&  2.3& 75.5& 11.0&  0.89$\pm$ 0.25& 4.53$\pm$ 0.59& 0.73$\pm$ 0.25& 0.90$\pm$ 0.44& 0.67$\pm$0.09&$-$0.72$\pm$0.09&0.11$\pm$0.29 &5 &11.8 &17.4 &2.2 &
\enddata
\tablenotetext{a}{The pointing ID (1--7) where the source is detected.}
\tablenotetext{b}{The offset angle in units of arcminutes from the mean optical axis in the corresponding pointing.}
\tablenotetext{c}{The total pn-equivalent exposure (sum of pn, MOS1, and
MOS2) at the source position in units of ksec, corrected for the
vignetting in the 0.5--4.5 keV band.}
\tablenotetext{d}{The background rate in the 0.5--4.5 keV band at the source position in units of $10^{-3}$ counts ksec$^{-1}$ arcsec$^{-2}$.}
\tablenotetext{e}{Set to be ``C'' if multiple-source fitting was
performed to obtain the fluxes to take into account possible source
confusion with nearby sources (see \S~3.2.3).}
\end{deluxetable}
\clearpage
\end{landscape}

\end{document}